\documentclass[twocolumn]{aa}
\usepackage{graphicx}
\usepackage{txfonts}
\usepackage{natbib}
\bibpunct{(}{)}{;}{a}{}{,} 
\newcommand{\ha}  {H$\alpha$}
\newcommand{\ew}  {EW(H$\alpha$)}

\def\simless{\mathbin{\lower 3pt\hbox
     {$\rlap{\raise 5pt\hbox{$\char'074$}}\mathchar"7218$}}}   
\def\simmore{\mathbin{\lower 3pt\hbox
     {$\rlap{\raise 5pt\hbox{$\char'076$}}\mathchar"7218$}}}   

\def\msun{~{\rm M}_\odot}
\def\rsun{~{\rm R}_\odot}

\begin{document}

   \title{Long-term optical variability of high-mass X-ray binaries. \\
   II. Spectroscopy}

   \subtitle{}
  \author{
        P. Reig\inst{1,2}
        \and
        A. Nersesian\inst{2}
        \and
        A. Zezas\inst{3,2,1}
        \and
        L. Gkouvelis\inst{4}
        \and
        M. J. Coe\inst{5}
          }

\authorrunning{Reig et al.}
\titlerunning{Spectroscopic variability of HMXBs}

   \offprints{pau@physics.uoc.gr}

   \institute{IESL, Foundation for Research and Technology-Hellas, 71110, 
                Heraklion, Greece \email{pau@physics.uoc.gr}
         \and Physics Department, University of Crete, 71003, 
                Heraklion, Greece 
        \and Harvard-Smithsonian Center for Astrophysics, 60 Garden Street, Cambridge, MA 02138, USA
         \and Observatorio Astron\'omico, Universidad de Valencia, 
         E-46100 Burjassot, Spain
         \and  Physics and Astronomy department, University of Southampton, Southampton SO17 1BJ, UK
        }

   \date{Received ; accepted}

\abstract
{High-mass X-ray binaries are bright X-ray sources. The high-energy
emission is caused by the accretion of matter from the massive companion onto a
neutron star.   The accreting material comes from either the strong stellar
wind in binaries with supergiant companions or the cirscumstellar disk in
Be/X-ray binaries. In either case, the \ha\ line stands out as the main source of
information about the state of the accreting material.
}
{We present the results of our monitoring program to study the  long-term
variability of the \ha\ line in high-mass X-ray binaries.  Our aim is to
characterise the optical variability timescales and study the interaction
between the neutron star and the accreting material.
}
{We  fitted the \ha\ line with Gaussian profiles and obtained the line
parameters and equivalent width. The peak separation in split profiles was
used to determine the disk velocity law and estimate the disk radius. 
The relative intensity of the two peaks (V/R ratio) allowed us to
investigate the distribution of gas particles in the disk. The equivalent
width  was used to characterise the degree of variability of the systems.
We  also studied the variability of the H$\alpha$ line in correlation with
the X-ray activity.
}
{Our results can be summarised as follows: 
i) we find that Be/X-ray binaries with narrow orbits are more variable than
systems with  long orbital periods,
ii) we show that a Keplerian  distribution of gas particles provides a
good description of the disks in Be/X-ray binaries,  as it does in
classical Be stars, 
iii) a decrease in the \ha\ equivalent width is generally observed after 
major X-ray outbursts, 
iv) we confirm that the \ha\ equivalent width correlates with disk radius, 
v) while systems with supergiant companions display multi-structured  
profiles, most of the Be/X-ray binaries show, at some epoch, double-peak asymmetric
profiles, which indicates that density inhomogeneities is a common property in 
the disk of Be/X-ray binaries, 
vi) the profile variability (V/R ratio) timescales are shorter and the \ha\ 
equivalent widths are smaller in Be/X-ray binaries than in isolated Be
stars, and 
vii) we provide new evidence that the disk in Be/X-ray binaries is, on average, 
denser than in classical Be stars.}
{We  carried out the most complete optical spectroscopic study of 
the global properties of high-mass X-ray binaries with the analysis of more than 
1 100 spectra from 20 sources. Our results provide further 
evidence for the truncation of the disk in Be/X-ray binaries. We conclude that 
the interaction between the compact object and the Be-type star works in two 
directions: the massive companion provides the source of matter for accretion,
affecting the surroundings of the compact object, and   the continuous
revolution of the neutron star around the optical  counterpart also produces the
truncation of the Be star's equatorial disk. }


   \maketitle

\begin{table*}
\caption{List of targets and relevant information.}
\label{targets}
\begin{center}
\begin{tabular}{@{}l@{~~~}l@{~~~}c@{~~~}c@{~~~}c@{~~~}c@{~~~}c@{~~~}c@{~~~}l@{~~~}l@{}}
\hline  \hline
X-ray             &Spectral &Mass$^\dag$ &Radius$^\dag$ &Orbital        &$e$    &$E(B-V)^{\dag\dag}$&Distance$^{\dag\dag}$  &disk-   &References\\
name              &type     &($\msun$)   &($\rsun$)     &period (d)     &        &(mag.)         &(kpc)          &loss    &for $P_{\rm orb}$ \& $e$\\
\hline
2S\,0114+65       &B1Ia         &16     &37     &11.60          &0.16   &1.33$\pm$0.04    &5.9$\pm$1.4  &--         &\citet{grundstrom07a}    \\
4U\,0115+63       &B0.2V        &17     &7.5    &24.32          &0.34   &1.71$\pm$0.05    &6.0$\pm$1.5  &yes        &\citet{raichur10}        \\
IGR\,J01363+6610  &B1V          &12.5   &6.3    &--              &--    &1.61$\pm$0.03    &2.2$\pm$0.5  &no         &---                      \\
RX\,J0146.9+6121  &B1V          &9.6    &5.4    &330?            &--    &0.88$\pm$0.03    &2.5$\pm$0.6  &no         &\citet{sarty09}          \\
IGR\,J01583+6713  &B2IV         &12.5   &6.3    &--             &--     &1.44$\pm$0.04    &3.4$\pm$0.8  &no         &---                      \\
RX\,J0240.4+6112  &B0.5V        &14.6   &6.9    &26.50          &0.54   &1.09$\pm$0.03    &1.6$\pm$0.4  &no         &\citet{aragona09}        \\ 
V\,0332+53        &O8.5V        &18.8   &7.9    &36.50          &0.42   &1.94$\pm$0.03    &6.0$\pm$1.5  &no         &\citet{raichur10}        \\
X\,Per            &O9.5III      &20.5   &13.2   &250            &0.11   &0.36$\pm$0.02    &0.9$\pm$0.4  &yes        &\citet{delgado01}        \\
RX\,J0440.9+4431  &B0.2V        &17     &7.5    &150*           &--     &0.91$\pm$0.03    &2.2$\pm$0.5  &yes        &\citet{ferrigno13}       \\
1A\,0535+262      &O9.7III      &20.5   &13.2   &111            &0.47   &0.77$\pm$0.04    &2.1$\pm$0.5  &yes        &\citet{janot87}          \\
IGR\,J06074+2205  &B0.5V        &14.6   &6.9    &--             &--     &0.86$\pm$0.03    &4.1$\pm$1.0  &yes        &---                      \\
AX\,J1845.0-0433  &O9Ia         &29.6   &21.8   &4.74           &0.34   &2.42$\pm$0.07    &5.5$\pm$1.5  &--         &\citet{gonzalez-galan15} \\
4U\,1907+09       &O9.5Iab      &27.8   &22.1   &8.36           &0.28   &3.31$\pm$0.10    &4.4$\pm$1.2  &--         &\citet{intzand98}        \\
XTE\,J1946+274    &B0--1IV-V    &15     &7      &172            &0.25   &1.18$\pm$0.04    &7.0$\pm$2.0  &no         &\citet{marcu15}          \\
KS\,1947+300      &B0V          &17.5   &7.7    &40.41          &0.03   &2.01$\pm$0.05    &8.0$\pm$2.0  &no         &\citet{galloway04}       \\
GRO\,J2058+42     &O9.5--B0V    &18     &8      &55*            &--     &1.37$\pm$0.03    &9.0$\pm$2.5  &no         &\citet{wilson05}         \\
SAX\,J2103.5+4545 &B0V          &17.5   &7.7    &12.67          &0.41   &1.36$\pm$0.03    &6.0$\pm$1.5  &yes        &\citet{baykal07}         \\
IGR\,J21343+4738  &B1IV         &12.5   &6.3    &--             &--     &0.75$\pm$0.03    &10.0$\pm$2.5 &yes        &---                      \\
4U\,2206+54       &O9.5V        &18     &8      &9.57           &0.30   &0.51$\pm$0.03    &3.0$\pm$0.7  &no         &\citet{stoyanov14}       \\
SAX\,J2239.3+6116 &B0V          &17.5   &7.7    &262*           &--     &1.66$\pm$0.04    &4.1$\pm$1.3  &no         &\citet{intzand01}        \\
\hline
\multicolumn{10}{l}{$\dag$: B star masses from \citet{porter96} and O star
masses from \citet{martins05}, except for 2S\,0114+65 \citep{reig96}.} \\
\multicolumn{10}{l}{$\dag\dag$: From \citet{reig15}, except for X Per
\citep{roche97}.} \\
\multicolumn{10}{l}{*: Based on the time interval between a small number of
type I outbursts and hence uncertain.} \\
\end{tabular}
\end{center}
\end{table*}

\section{Introduction}

High-mass X-ray binaries (HMXB) are accretion-powered binary
systems where a neutron star orbits an early-type (O or B) companion. The
luminosity class of the optical companion subdivides HMXBs into Be/X-ray
binaries (BeXB), when the optical star is a dwarf, subgiant, or giant OBe
star (luminosity class III, IV, or V) and supergiant X-ray binaries (SGXBs),
when they contain an evolved star of luminosity class I-II. In
SGXBs, the donor produces a  substantial  stellar  wind, removing 
between  $10^{-6}-10^{-8}$  M$_{\odot}$  yr$^{-1}$  with terminal
velocities up to 2000 km s$^{-1}$.  A neutron star in a relatively  close
orbit will capture a  significant  fraction of this wind, sufficient to
power a bright X-ray source. In BeXB, the donor is a Be star, i.e. a
rapidly rotating star of spectral type early-B or late-O with a gaseous
equatorial disk \citep{ziolkowski02,negueruela07a,paul11,reig11}. 

The disk is the main source of variability in BeXBs because it evolves on
much faster timescales than other components of the binary. On one hand,
the disk itself emits at optical and infrared wavelengths, contaminating
the continuum emission intensity (magnitudes and colours) and the spectral
lines of the underlying star, making it difficult to determine  
astrophysical parameters  \citep{fabregat98,riquelme12}. On the other hand,
the disk is also responsible for the X-ray variability because it constitutes the
source of matter available for accretion. Thus the study of the evolution
of the disk enables us to characterise the variability timescales at
different wavelengths.  

We have been monitoring the HMXBs visible from the northern hemisphere in
the optical band since 1999. The monitoring consists of BVRI and
medium-resolution spectra around the \ha\ line. JHK photometry has been
performed since 2009 and optical polarimetry in the $R$ band has been
regularly obtained since 2013. The results from the photometric study were
presented in \cite{reig15}. In this work, we present the results of the
spectroscopic observations. Although we also studied the long-term spectral
emission of SGXBs, we primarily focus on BeXBs because of their larger
amplitude of variability on long timescales.  

In a study of the long-term photometric variability of BeXB, \citet{reig15}
find that when timescales of years are considered, a good correlation
between the X-ray and optical variability is observed. However, photometry
gives us information on the optical continuum, which is formed in the
innermost part of the disk.  The \ha\ line stands out as a better proxy for
the state of the disk \citep{quirrenbach97,tycner05,grundstrom06} because
it is formed further away from the star in a larger part
of the disk. Moreover, the outer parts of the disk are expected to be  more
strongly affected by the gravitational pull of the neutron star companion.
Thus, we would, in principle, expect stronger and faster changes in the
parameters of the line, which would reflect a  major disruption of the
disk.  Large amplitude changes in the strength and profile shape of the
\ha\ line have been linked to structural changes of the circumstellar
decretion disk \citep{roche93b,haigh04,reig07b,moritani13,reig14a}.

The main objective of this work is to study the patterns of variability of
BeXB as a group with emphasis on two questions. One relates to the
long-term variability of  the spectral parameters of the \ha\ line and 
their relationship with the X-ray activity of the system. Previous works on
correlated optical/X-ray variability focused on individual systems, which
did no allow   general conclusions to be drawn on disk variability
timescales and the correlation with X-ray emission in BeXB as a population.
The other question relates to the search for further observational evidence
of the interaction between the neutron star and the disk. Most of our
understanding of the properties of the equatorial disk stems from the study
of classical Be stars, i.e. isolated systems without a neutron star
companion. However, the conditions in classical Be stars may not be the
same as in BeXB. For example, there is growing evidence that the disk in
BeXBs is truncated by the neutron star
\citep{reig97a,zamanov01,negueruela99,okaneg01,okazaki02,reig11}. Being
able to understand  this interaction could provide new insights into the
accretion mechanism in X-ray pulsars.

\begin{figure*}[t]
\begin{center}
\includegraphics[width=18cm]{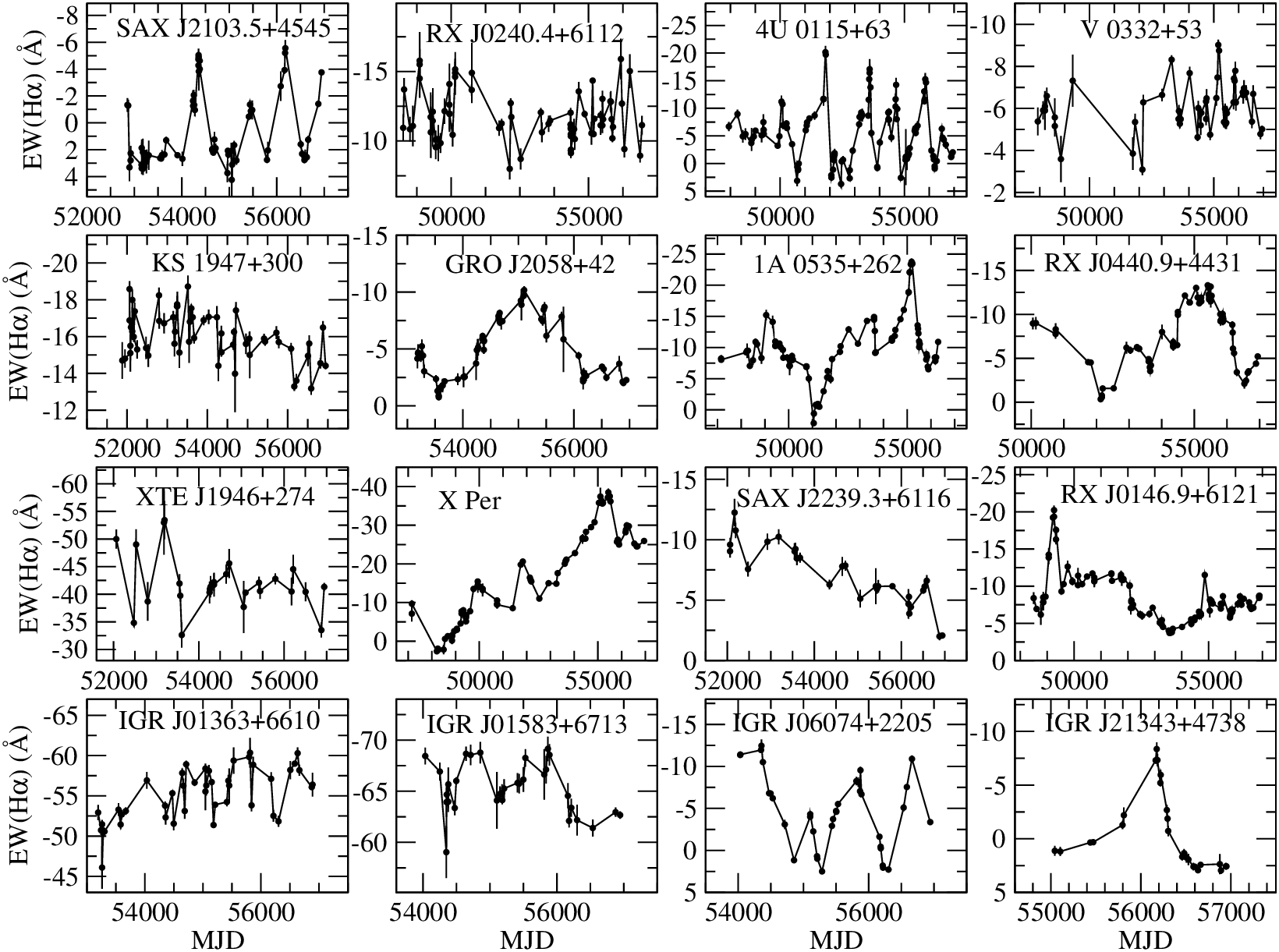} 
\end{center}
\caption[]{Evolution of the \ha\ equivalent width of the BeXBs
considered in this work, ordered by increasing value of the orbital
period (see Table~\ref{targets}). The orbital period of the four sources at the bottom
is not known. }
\label{diskew}
\end{figure*}

\section{Observations and data reduction}
\label{observations}

We have been monitoring the \ha\ line of HMXBs visible from the northern
hemisphere using the 1.3 m telescope of the Skinakas Observatory (SKO)
located in the island of Crete (Greece) since 1999. The targets have also
been observed regularly using the 1.5 m telescope of the Fred Lawrence
Whipple Observatory (FLWO) at Mt. Hopkins (Arizona) since 2007. The
instrumental set-up for the 1.3\,m telescope of the Skinakas Observatory
was a 2000$\times$800 ISA SITe CCD and a 1302 l~mm$^{-1}$ grating, giving a
nominal dispersion of $\sim$1.04 \AA/pixel, while for the 1.5-m telescope at
Mt. Hopkins (Arizona),  the FAST-II spectrograph \citep{fabricant98} plus
FAST3 CCD, a backside-illuminated 2688x512 UA STA520A chip with 15$\mu$m
pixels and a 1200 l~mm$^{-1}$ grating (0.38 \AA/pixel). Occasionally,  a
2400 l~mm$^{-1}$ and a 600 l~mm$^{-1}$ grating was used in Skinakas and Mt.
Hopkins observations, respectively. Additionally, we also used data
obtained in service mode  with the 4.2-m William Herschel Telescope  (WHT)
and the 2.5m Nordic Optical Telescope (NOT) in El Roque de los Muchachos
observatory in La Palma (Canary Islands, Spain). The spectra were reduced
with the dedicated packages for spectroscopy of the {\tt STARLINK} or 
{\tt IRAF} projects following the standard procedure. The images were bias-subtracted and flat-field corrected. Spectra of comparison lamps were taken
before each exposure  to account for small variations of the
wavelength calibration during the night. Finally, the spectra were
extracted from an aperture that encompassed more than 90\% of the flux of the
object. Sky subtraction was performed by measuring the sky spectrum from an
adjacent object-free region. For the FLWO spectra, this procedure is carried
out with the RoadRunner reduction system \citep{tokarz97}.

Some of the observations have been published in papers dedicated to
individual sources \citep{reig05b,blay06,coe06,reig07b,reig10b,reig14a}.
For ten sources from the target list, we complemented our observations with
data taken before 1999. These spectra were obtained in the framework of the
Southampton/Valencia collaborative project \citep{reig97c}, where a number
of telescopes were used: the 2.5m Isaac Newton Telescope (INT) and the 1.0m
Jacobus Kapteyn Telescope (JKT), both located at the Roque de los Muchachos
Observatory in La Palma, the 1.5m at Palomar Mountain (PAL), the 1.9 m
telescope of the South Africa Astronomical Observatory (SAAO), and the 2.2m
telescope at Calar Alto  in Almer\'{\i}a, Spain (CA), and resulted in
numerous publications 
\citep{clark98,clark99,clark01,coe93,coe96,gorrod93,haigh99,haigh04,negueruela97,negueruela98a,negueruela99,negueruela01b,norton91,reig96,reig97b,reig00,roche93a,roche93b,unger98}.
For the sake of homogeneity, we re-analysed all the spectra in a consistent
way, as explained in the next section.  The list of targets is given in
Table~\ref{targets}.

\begin{table}
\caption{Average, standard deviation, minimum, and maximum values of the \ha\ 
equivalent width. Also indicated is the number of spectra analysed. 
Negative values of the equivalent width mean emission profiles.}
\label{minmax}
\begin{center}
\begin{tabular}{@{}l@{~~~~}c@{~~~~}c@{~~~~}c@{~~~~}c@{~~~~}c@{}}
\hline  \hline
Source           &Average       &Stand. &Min.   &Max.   &N\\     
                &(\AA)          &(\AA)  &(\AA)  &(\AA)  & \\
\hline
2S\,0114+65      &--1.8  &0.5 &--0.4  &--3.2   &57    \\
4U\,0115+63      &--5.6  &5.2 &+3.7   &--20.2  &106   \\
IGR\,J01363+6610 &--55.4 &3.6 &--46.1 &--59.8  &45    \\
IGR\,J01583+6713 &--65.2 &2.5 &--59.0 &--69.1  &32    \\
RX\,J0146.9+6121 &--8.4  &3.3 &--3.7  &--20.2  &106   \\
RX\,J0240.4+6112 &--11.7 &1.8 &--8.0  &--15.9  &71    \\
V\,0332+53       &--6.0  &1.2 &--3.1  &--9.0   &53    \\
X Per            &--17.0 &11.9 &+2.6  &--38.5  &76    \\
RX\,J0440.9+4431 &--7.5  &3.6 &--0.3  &--13.3  &65    \\
1A\,0535+262     &--9.9  &5.5 &+2.1   &--23.8  &87    \\
IGR\,J06074+2205 &--4.7  &4.5 &+2.5   &--12.5  &36    \\
AX\,J1845.0-0433 &--3.8  &0.6 &--2.6  &--5.0   &29    \\ 
4U\,1907+09      &--6.8  &1.5 &--3.2  &--10.7  &48    \\
XTE\,J1946+274   &--40.8 &4.1 &--32.6 &--49.9  &26    \\
KS\,1947+300     &--16.2 &1.3 &--13.2 &--18.7  &51    \\
GRO\,J2058+42    &--4.7  &2.6 &--0.8  &--10.2  &57    \\
SAX\,J2103.5+4545&+0.7   &2.7 &+4.2   &--5.6   &75    \\
IGR\,J21343+4738 &--0.5  &3.6 &+3.0   &--8.4   &26    \\
4U\,2206+54      &--3.1  &1.0 &--1.1  &--5.1   &90    \\
SAX\,J2239.3+6116&--7.0  &2.5 &--2.0  &--12.3  &28    \\
\hline
\end{tabular}
\end{center}
\end{table}

\section{Data analysis}

The spectral analysis was performed with the ~{\it splot} task in {\tt
IRAF}. To ensure homogeneous processing, the spectra were normalized with
respect to the local continuum.  The definition of the continuum level is
crucial because it represents the main source of uncertainty in the
determination of the spectral parameters. To normalize the continuum, we
divided the spectrum by a smooth curve that was obtained by fitting regions
of the continuum devoid of spectral features on both sides of the line.
This process sets the continuum to unity, and allows us to measure the
lines from spectrum to spectrum in a consistent way. Ten different
combinations of spectral regions and fitting functions provided ten
measurements of the continuum level and hence ten independent measurements
of the spectral parameters. The final value and error were taken to be the
average and standard deviation of those ten measurements for each
parameter.

The \ha\ line exhibited a wide range of profiles, both symmetric and
asymmetric. Symmetric profiles include single- and double-peaked shapes. In
some cases, the central depression between the peaks is sharp and deep,
going beyond the continuum. This profile is known as a shell profile
and is thought to arise when the observer's line of sight toward the
central star intersects  parts of the disc, which is cooler than the
stellar photosphere \citep{hanuschik95,rivinius06}. Absorption lines were
also observed in a number of sources. 

To extract the line parameters, we fit these
profiles with one, two, or three Gaussian functions, depending on whether the
line showed a single, split, or shell profile (in this case the Gaussian
that corresponds to the central depression had opposite sign to the other two).

The fits provided the following parameters: line centres, that is,  the
wavelength at which the intensity of the line is maximum, full-width at
half maximum (FWHM), and peak intensity above the normalised continuum. In
double-peak lines, the peak at shorter wavelength is referred to as the
blue or violet peak, while the peak at longer wavelength is
known as the red peak. 

Using the best-fit parameters, we obtain the two following  quantities:

\begin{itemize}

\item[-] peak separation ($\Delta V$), defined as the difference between
the central wavelengths of the red and the blue peak in velocity units
$\Delta V= \Delta \lambda/\lambda_0 \times c$, where $c$ is the speed of
light, and $\lambda_0$ is the wavelength of the \ha\ line, 6562.8 \AA.

\item[-] $V/R$ ratio, defined as the ratio of the relative intensity at the
blue and red emission peak maxima (after subtracting the underlying
continuum). For plotting purposes, we used the log of this ratio,
$\log(V/R)$. Thus negative values indicate a red-dominated peak ($V < R$),
positive values a blue-dominated profile ($V > R$), and values close to
zero correspond to equal intensity peaks ($V\approx R$).

\end{itemize}

In addition, the equivalent width (\ew) of the entire line was also
calculated directly from the data, after normalisation. The \ew\ of each
individual observation is given in the appendix.

\begin{figure}
\begin{center}
\includegraphics[width=8cm]{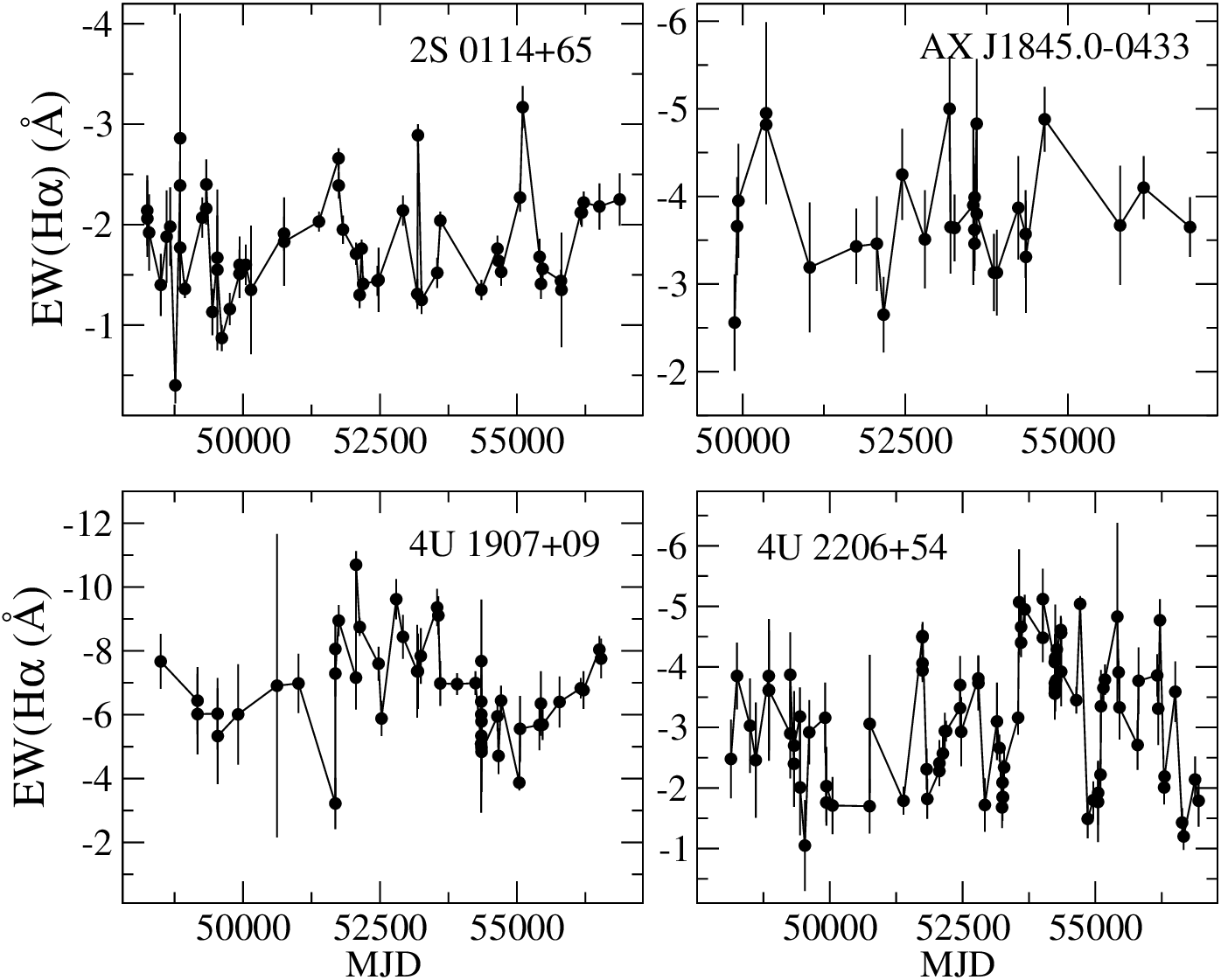}
\end{center}
\caption[]{Evolution of the \ha\ equivalent width of wind-fed sources. }
\label{wind}
\end{figure}

\section{Results}

One of the most useful observables that can be used to monitor the spectral
variability in HMXBs is the \ha\ line. Both SGXBs and BeXBs show this line
in emission because they present a significant amount of circumstellar
material. The \ha\ line is formed by recombination either in a large part
of the equatorial disk, in the case of BeXBs, or in the low-density,
high-velocity spherical wind in SGXBs.  Other spectral emission lines such
as those of helium and iron may also originate in the equatorial disk.
However,  they either probe only the innermost regions of the disk (He I
lines) or are much more difficult to detect because they are narrower and
weaker (Fe lines) than hydrogen lines \citep{hanuschik87}. In addition, the
strength of the \ha\ line is a good indicator of the size of the equatorial
disk. This has become evident thanks to interferometric observations that
resolved the disk and allowed the measurement of the angular dimensions of
Be star disks. \citet{quirrenbach97} and \citet{tycner05} showed that there
is a clear dependence of the net \ha\ emission on the physical extent of
the emitting region. \citet{grundstrom06} demonstrated that there are
monotonic relationships between the emission-line equivalent widths and the
ratio of the angular half-width at half-maximum of the projected disk's
major axis to the radius of the star. In addition to the strength of the
\ha\ line, the profile shape is also important to study the conditions in
the disk. Asymmetries in the line profile imply a distorted disk due to
warping or a perturbed disk owing to density anisotropies
\citep{okazaki91,reig00,negueruela01b,okazaki13,martin14}.

In this section we study the variability of the strength (equivalent width)
and shape (line profile) of the \ha\ line on timescales of months to years,
and the impact that the neutron star has on it. We compare the distribution
of \ew\ in isolated Be stars and BeXB systems and study the variation of
the maximum value of \ew\ as a function of orbital separation. We also
investigate the disk rotational velocity law based on the peak separation
of the split profile of the \ha\ line and find that it is consistent with
Keplerian rotation. The peak separation can also be used to estimate the
radius of the disk \citep{andrillat90,jaschek92,dachs92}. We provide the
first observational correlation between the disk radius and \ew\ in BeXBs.
Finally, we compare the evolution of the \ew\ with the X-ray variability
and show that the disk is strongly disrupted after major X-ray outbursts.

\begin{table*}
\caption{Results from the variability analysis. The smaller the p-value,
the more variable the system is.  }
\label{anova}
\begin{center}
\begin{tabular}{l|ccccc|ccccc}
\hline  \hline
Source          &\multicolumn{5}{c}{EW(H$\alpha$)}       &\multicolumn{5}{|c}{$F_V$ }  \\
                &$rms$ (\%)     &$s/\bar{x}$ (\%)       &$F$-value      &$p$-value      &$N$
                &$rms$ (\%)     &$s/\bar{x}$ (\%)       &$F$-value      &$p$-value      &$N$    \\       
\hline
2S\,0114+65      &22 &28 &  2.6 &$2.2\times10^{-4}$  & 57  &--  & 4 & 0.2 &$9.9\times10^{-1}$ &11 \\
4U\,0115+63      &91 &92 & 29.4 &$4.4\times10^{-49}$ &106  &20  &21 &11.6 &$1.0\times10^{-8}$ &27 \\
IGR\,J01363+6610 & 6 & 6 &  7.8 &$1.5\times10^{-10}$ & 45  &--  & 5 & 1.0 &$5.1\times10^{-1}$ &17 \\
IGR\,J01583+6713 & 3 & 3 &  4.0 &$1.0\times10^{-4}$  & 32  &--  & 3 & 0.3 &$9.9\times10^{-1}$ &17 \\
RX\,J0146.9+6121 &39 &39 & 47.5 &$1.7\times10^{-59}$ &106  &--  & 4 & 0.3 &$9.9\times10^{-1}$ &26 \\
RX\,J0240.4+6112 &14 &15 &  3.9 &$2.7\times10^{-8}$  & 71  &--  & 4 & 0.3 &$9.6\times10^{-1}$ &11 \\
V\,0332+53       &19 &21 &  6.3 &$2.9\times10^{-10}$ & 53  &8   & 9 & 5.1 &$1.6\times10^{-2}$ & 9 \\
X\,Per           &70 &70 &270.0 &$8.9\times10^{-71}$ & 76  &--  &-- &--   &--                   &-- \\
RX\,J0440.9+4431 &48 &48 & 51.0 &$2.6\times10^{-37}$ & 65  &7   & 9 & 3.3 &$1.6\times10^{-2}$ &15 \\
1A\,0535+262     &55 &55 &144.7 &$2.3\times10^{-69}$ & 87  &5   &11 & 1.3 &$3.7\times10^{-1}$ & 9 \\
IGR\,J06074+2205 &94 &94 &191.4 &$2.3\times10^{-31}$ & 36  &4   &11 & 1.2 &$4.5\times10^{-1}$ & 5 \\
AX\,J1845.0-0433 &7 &17 &  1.2 &$3.0\times10^{-1}$  & 29  &--   & 3 & 0.3 &$9.5\times10^{-1}$ &11 \\ 
4U\,1907+09      &13 &21 &  1.7 &$2.9\times10^{-2}$  & 48  &--  & 4 & 0.6 &$8.6\times10^{-1}$ &14 \\
XTE\,J1946+274   & 8 &10 &  2.8 &$5.6\times10^{-3}$  & 26  &--  & 6 & 0.6 &$7.3\times10^{-1}$ & 8 \\
KS\,1947+300     & 5 & 8 &  1.8 &$2.4\times10^{-2}$  & 51  &--  & 2 & 0.1 &$9.9\times10^{-1}$ &18 \\
GRO\,J2058+42    &53 &54 &15.5  &$3.1\times10^{-20}$ & 57  &5  & 7 & 2.3 &$3.3\times10^{-2}$ &21 \\
SAX\,J2103.5+4545&370  &380  &22.0 &$3.3\times10^{-31}$ & 75  &18 &19 &15.0 &$1.0\times10^{-6}$ &17\\
IGR\,J21343+4738 &763  &757  &67.0 &$2.8\times10^{-17}$ & 26  &6 &7 &2.4 &$1.8\times10^{-1}$ &6\\
4U\,2206+54      &29 &33 & 4.0 &$1.3\times10^{-10}$ & 90  &--   & 4 & 0.2 &$9.9\times10^{-1}$ &28\\
SAX\,J2239.3+6116&35 &36 &15.8 &$1.5\times10^{-10}$ & 28  &--   & 7 & 0.7 &$6.5\times10^{-1}$ & 7\\
\hline
\end{tabular}
\end{center}
\end{table*}

\subsection{Measuring variability}

In this section we perform a variability analysis of the equivalent width
of the \ha\ line. Figures~\ref{diskew} and \ref{wind} show the evolution of
\ew\ for the BeXBs and the SGXBs that were analysed in this work, respectively, and
Table~\ref{minmax} gives the mean, standard deviation, largest and smallest
values of \ew,\ and number of observations for each source. To quantify the
degree of variability, we used the root-mean-square amplitude defined as 

\begin{equation}
rms=\sqrt{\sigma^2_{\rm NXS}}
,\end{equation}

\noindent where $\sigma_{\rm NXS}$ is the normalized excess
variance \citep[see e.g.][]{vaughan03}

\begin{equation}
\sigma^{2}_{\rm NXS}=\frac{s^2-\overline{\sigma_{\rm err}^2}}{\bar{x}^2}
,\end{equation}

\noindent $\bar{x}$ is the mean value and $s^2$ is the observation (or sample) variance, i.e. the square of
the standard deviation 

\begin{equation}
s^2=\frac{\sum\limits_{i=1}^{N}(x_i-\bar{x})^2}{N-1}
,\end{equation}

\noindent and $\overline{\sigma_{\rm err}^{2}}$ is the variance expected
from the errors of the measurements

\begin{equation}
\overline{\sigma_{\rm err}^2}=\frac{\sum\limits_{i=1}^{N}{\sigma_i^2}}{N}
,\end{equation}

\noindent where $N$ is the number of measurements and $\sigma_i$ the error
of each measurement. Because, in some cases, the excess variance was
negative, we also computed the ratio 

\begin{equation}
\label{r-par}
r=\frac{s}{\bar{x}}
\end{equation}

\noindent of the standard deviation $s$ over the mean value $\bar{x}$. Here
the variable $x$ refers to the equivalent width of the \ha\ line.

From the definition of excess variance, it is clear that negative values
mean that the source is not variable. Negative values of the excess
variance imply that the variance based on the entire set of observations
$s^2$ (sample variance) is smaller than the variance based on the 
uncertainty of the individual measurements $\overline{\sigma_{\rm err}^2}$.
In other words, the dispersion of the observations is smaller than the mean
individual errors.

Finally,  we run an $F$ test and computed the probability $p$-value
considering the $F$-value, $F=s^2/\overline{\sigma_{\rm err}^2}$. The null
hypothesis is that the two variances are equal. In our case, the rejection
of the null hypothesis (low $p$-value) means that the source is variable. A
similar analysis has been used by \citet{jones11} for a sample of isolated
Be stars\footnote{Note that we use the average of the square of the errors
while \citet{jones11} calculated the square of the average error as the
internal variance.}.

The optical continuum and line emission probe different parts of the disk.
The \ha\ line is formed in the outer parts of the disk, while continuum
emission originates in the inner parts. Thus to have a more complete picture of
the disk variability, we also perform the statistical analysis  on the flux
of the $V$ band given by

\begin{equation}
F_V=10^{-0.4\,(V-A_V+21.10)}
.\end{equation}

\noindent Although the $V$ band has a contribution from both the star and 
the disk, the long-term changes are expected to have their origin in the
disk. For information about the photometric observations and data, see \citet{reig15}. The constant $m_z(V)=21.10=-2.5\log
(3.631\times 10^{-9})$ is the zero-point magnitude for the $V$ band in erg
s$^{-1}$ cm$^{-2}$ \AA$^{-1}$ \citep{bessell98}.
Table~\ref{anova} summarises the results of our statistical analysis.

\begin{figure}
\begin{center}
\includegraphics[width=8cm]{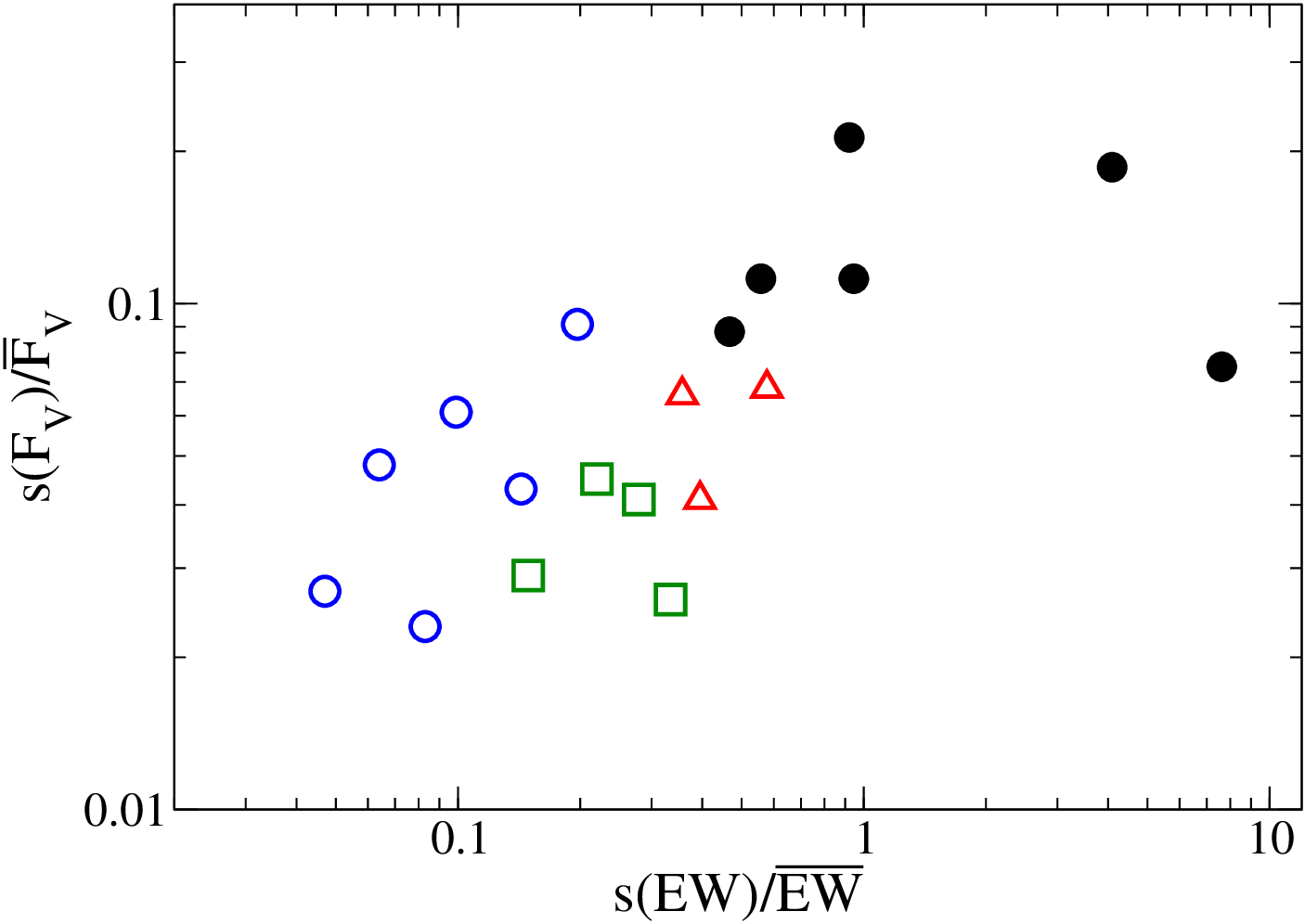}
\end{center}
\caption[]{Fractional amplitude of variability of the optical continuum (V
band) and \ha\ line. Black filled circles: BeXBs that have gone through a
disk-loss phase; blue empty circles: BeXBs with stable disks; red triangles:
other BeXBs; green empty squares: wind-fed accreting systems. }
\label{rms-ew-v}
\end{figure}
\begin{figure}
\begin{center}
\includegraphics[width=8cm]{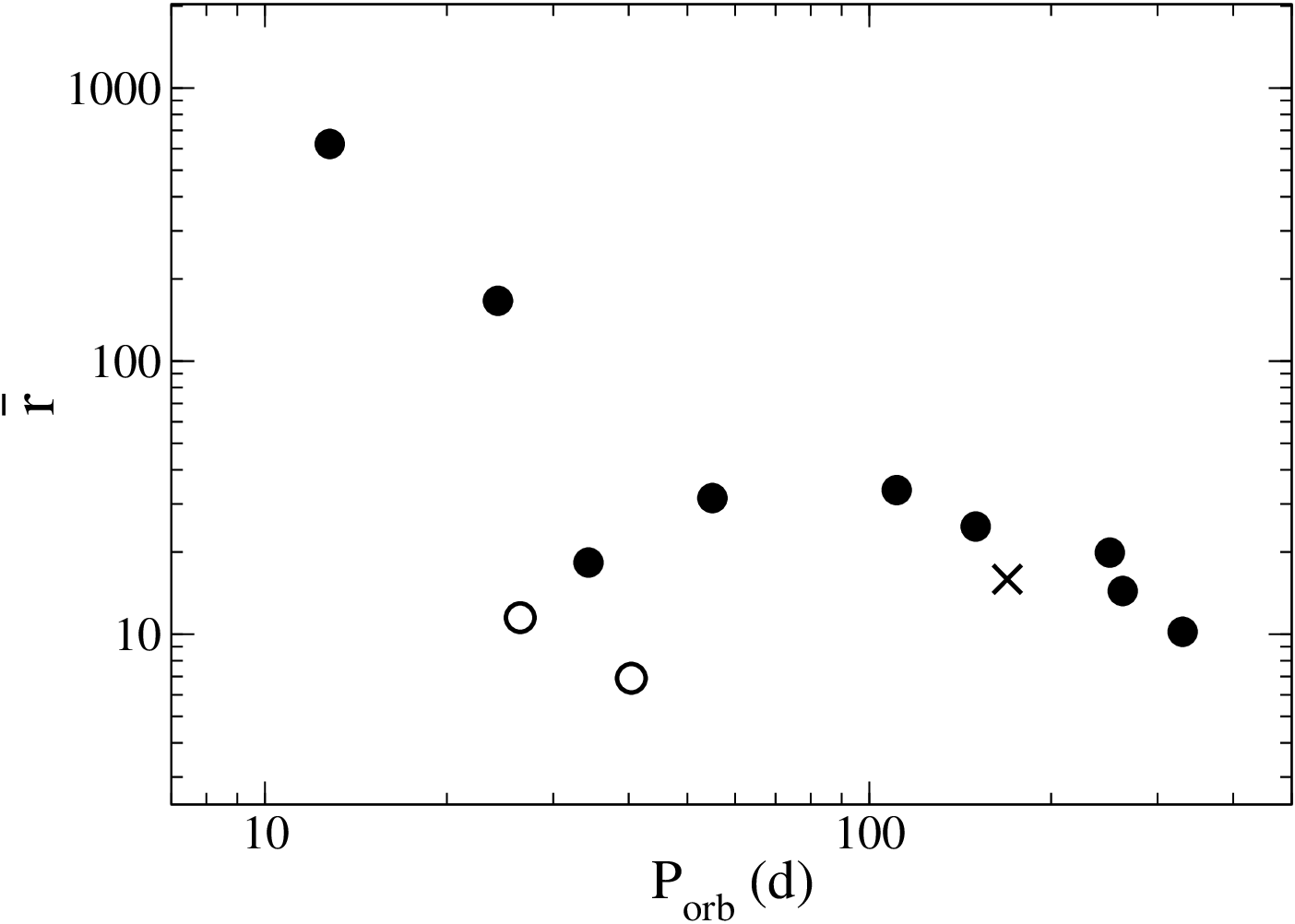}
\end{center}
\caption[]{\ha\ equivalent width variability as a function of the orbital period 
in BeXBs with orbital solutions. The Y-axis is the average ratio of the standard 
deviation over the mean calculated in intervals of 500 days (or 1000 days for
XTE\,J1946+274, cross). Empty circles correspond to peculiar systems (see text).}
\label{rms-orb}
\end{figure}

\subsection{Variability diagram}
\label{varcor}

Figure.~\ref{rms-ew-v} shows the relationship between the variability in the
continuum and the line. A number of facts are worth noting: 

\begin{itemize}

\item[--]  Seven binaries exhibited \ha\ in absorption (\ew\ $\simmore$ 0)
at some moment during the period covered by our observations
(Table~\ref{targets} and Fig.~\ref{diskew}): 4U\,0115+63,
X\,Per\footnote{X\,Per does not have photometric information because it is
too bright for the SKO 1.3 m telescope.},
RX\,J0440.9+4431\footnote{Although a full  absorption profile was not
observed in the \ha\ line of RX\,J0440.9+4431, a disk-loss episode is
likely to have occurred in early 2001, when there is a gap of about ten
months in the observations \citep[see discussion in][]{reig05b}. In
Fig.~\ref{rms-ew-v}, this source is the filled black circle that lies
closer to the triangle symbols.}, 1A\,0535+262, IGR\,J06074+2205,
SAX\,J2103.5+4545, and IGR\,J21343+4738. This is generally interpreted as
the loss of the disk. BeXBs that go through disk-loss episodes (filled
black circles in Fig.~\ref{rms-ew-v}) occupy the upper right part of the
diagram, indicating that  these systems are the most variable, both
spectroscopically and photometrically.

\item[--] On the opposite side of the diagram lie the systems with very
stable disks (blue empty circles in Fig.~\ref{rms-ew-v}). These are systems
that do not show significant optical spectral variability in the period
covered by the observations. They tend to have very large disks (i.e. large
\ew). IGR J01363+6610, IGR J01583+6713, XTE J1946+274, and KS 1947+300  belong
to this category. These systems do not exhibit
significant changes in the spectral line parameters. They normally show
symmetric single-peaked profiles that remain unchanged for a long period of
time. However, they exhibit moderate photometric variability.

\item[--] In between these two extreme cases (red empty triangles in
Fig.~\ref{rms-ew-v}), we find BeXBs displaying structural changes in the
disk (V/R variability, long-term weakening) but without losing the disk.
Although some systems nearly revert emission into absorption, e.g.
RX\,J0146.9+6121, GRO\,J2058+42, or SAX\,J2239.3+6116, \ew\ never
goes below zero during the time covered by our observations. 

\item[--]  On average, systems for which X-rays results from accretion from
a stellar wind (the SGXBs 2S\,0114+65, AX\,J1845.0-0433, 4U\,1907+09, and
also 4U\,2206+54, green empty squares) display smaller amplitude of
variability in \ew\ (and brightness) than BeXB systems, although
spectroscopically they can be more variable than BeXBs with stable
disks\footnote{We note that the comparison is done on the amplitude of
variability relative to the mean value and not changes in absolute values.
The absolute changes of \ew\ in wind-fed systems are considerably smaller
than in BeXBs.} (Fig.~\ref{wind}). This result implies that, in a stable
configuration where the Be star feeds the disk constantly, the optical
emission from this type of rotationally supported mass distribution is more stable
than from the inhomogeneous fast varying spherical stellar wind. 

\item[--] \ew\ varies over a larger fractional range than the $V$
magnitude. This can be understood taking into account the different
formation loci of continuum and line emission. Non-LTE Monte Carlo
radiative transfer simulations \citep{carciofi06} show that the emission in
the $V$ band is formed in the innermost parts of the disk, very close to
the star, whereas \ha\ emission forms further away from the star. For
example, in the case of a rapidly rotating B1Ve star, typical of BeXBs, the
simulation showed that approximately 95\% of the continuum excess comes
from within 2 $R_*$ whereas, for the \ha\ line, the disk emission only fills
the photospheric profile when the disk is about 5$R_*$ \citep{carciofi11}.
Figure~\ref{rms-ew-v} confirms the idea that the external parts of the disk
are more prone to changes that the inner parts.

\end{itemize}

\subsection{Variability versus orbital period}
\label{varporb}

Next we investigated whether the neutron star affects the variability
patterns. The idea is simple: one would expect that the closer the two
components of the binary are, the stronger the tidal torque exerted on the disk by the
neutron star.

In Fig.~\ref{diskew}, we have ordered the systems in increasing value of the
orbital period (see Table~\ref{targets}).  Four systems (IGR\,J01363+6610,
IGR\,J01583+6713, IGR\,J06074+2205, and IGR\,J21343+4738) do not have a
known orbital period.  Visual inspection of Fig.~\ref{diskew} reveals
that systems with short orbital periods ($P_{\rm orb} < 50$ d) exhibit fast
variability, i.e. choppy curves, while systems with longer orbital period
show slow changes, i.e. smooth curves. Systems with short orbital periods
do not show long-term increasing or decreasing trends. Instead, \ew\
`oscillates' around a certain mean value. 

We quantified this result in the following way: we divided the long-term
\ew\ curve  into 500-d intervals. For each interval, we computed the ratio
of the standard deviation over the mean, $r=s/\overline{EW(H\alpha)}$.
Finally, we obtain the average $\bar r$ for each source. In this
calculation we used intervals with at least four measurements of \ew.
Figure~\ref{rms-orb} shows $\bar r$ as a function of the orbital period for
all the BeXB whose orbital period is known (see Table~\ref{targets}).  One
can observe a distinct trend with the narrower orbit systems showing  more
variability. A simple explanation is that wider orbit systems are able to
develop stable disks during long periods of time. In contrast, the systems
with short orbital periods feel the tidal truncation exerted by the neutron
star more strongly and more frequently, so that the disk does not easily
achieve a stable configuration.

The choice of the 500-d duration interval is somewhat arbitrary
but it is justified by the need to sample the fast variations of the
shorter period systems. We note that for large intervals ($\simmore 1000$ d)
the amplitude of variations of long-period systems begins to be comparable
to the shorter period systems. We also note that although the
value of $\bar r$ for the individual systems may vary, the decreasing trend of
Fig.~\ref{rms-orb} remains for any choice of the duration of the interval
in the range 300--1000 days. The cross symbol in Fig.~\ref{rms-orb}
corresponds to  XTE\,J1946+274 and simply denotes the fact that owing to
large observational gaps, the duration of the interval in
this source was taken to be 1000 d instead of 500 d, since the shorter
duration did not provide enough number of intervals with more than 4 points.

There are three systems that deviate from the general trend: RX\,J0240.4+6112
and KS\,1947+300 (empty circles in Fig.~\ref{rms-orb}) and to a lesser
extent V\,0332+53.
RX\,J0240.4+6112 is not a typical BeXB. The nature of the compact object
(neutron star or black hole) is not known and no X-ray pulsations have been
detected so far. It belongs to the class of $\gamma$-ray binaries
\citep{dubus13} and is the only system of our list that exhibits radio
emission, which is associated either with a jet or with the interaction between the
relativistic wind of a young non-accreting pulsar and the wind of the donor
star. Although the optical counterpart is a B0Ve, the physical conditions
that prevail in the disk are unknown.

KS\,1947+300 is the only BeXB of our sample with a near circular orbit
\citep{galloway04}. According to the viscous disk truncation model, these
systems are truncated at the 3:1 resonance radius and the gap size between
the truncation radius and the radius where the gravity by the neutron star
begins to dominate (critical lobe radius) is considerably wider than in
systems with larger eccentricities. Wide gaps result in disk truncation
that is very effective and the disk can accumulate mass over a long period of
time \citep{okaneg01}. Assuming typical masses for a B0Ve for the optical
and compact components of the binary, the orbital parameters  
\citep[$P_{\rm orb}=40.4$ d and $e=0.03$,][]{galloway04}, the viscous
decretion disk model predicts a gap size of $\Delta r/a\sim0.24$, where $a$
is the binary orbital separation, which is significantly larger than the
values found for systems with highly-eccentric orbits \citep[see Table 2
in][]{okaneg01} for which $\Delta r/a\simless0.1$, typically.  Therefore, we
conclude that although the neutron star effectively acts as a barrier that
prevents the free expansion of the disk, the Be star in KS\,1947+300 can
reach a stable configuration owing to the large gap between the disk and
the neutron star.

The case of V\,0332+53 is harder to explain. Of the BeXBs with reliable
orbital solutions, it is the only system with a periastron distance of the
order of, or smaller than, 10 stellar radii that shows little variability.
One possible explanation could be the low inclination angle
($\simless10^{\circ}$) of the orbit \citep{negueruela99}, i.e. the disk is
seen pole-on, hence the entire disk is exposed to the observer all the
time.

\begin{figure}
\begin{center}
\includegraphics[width=8cm]{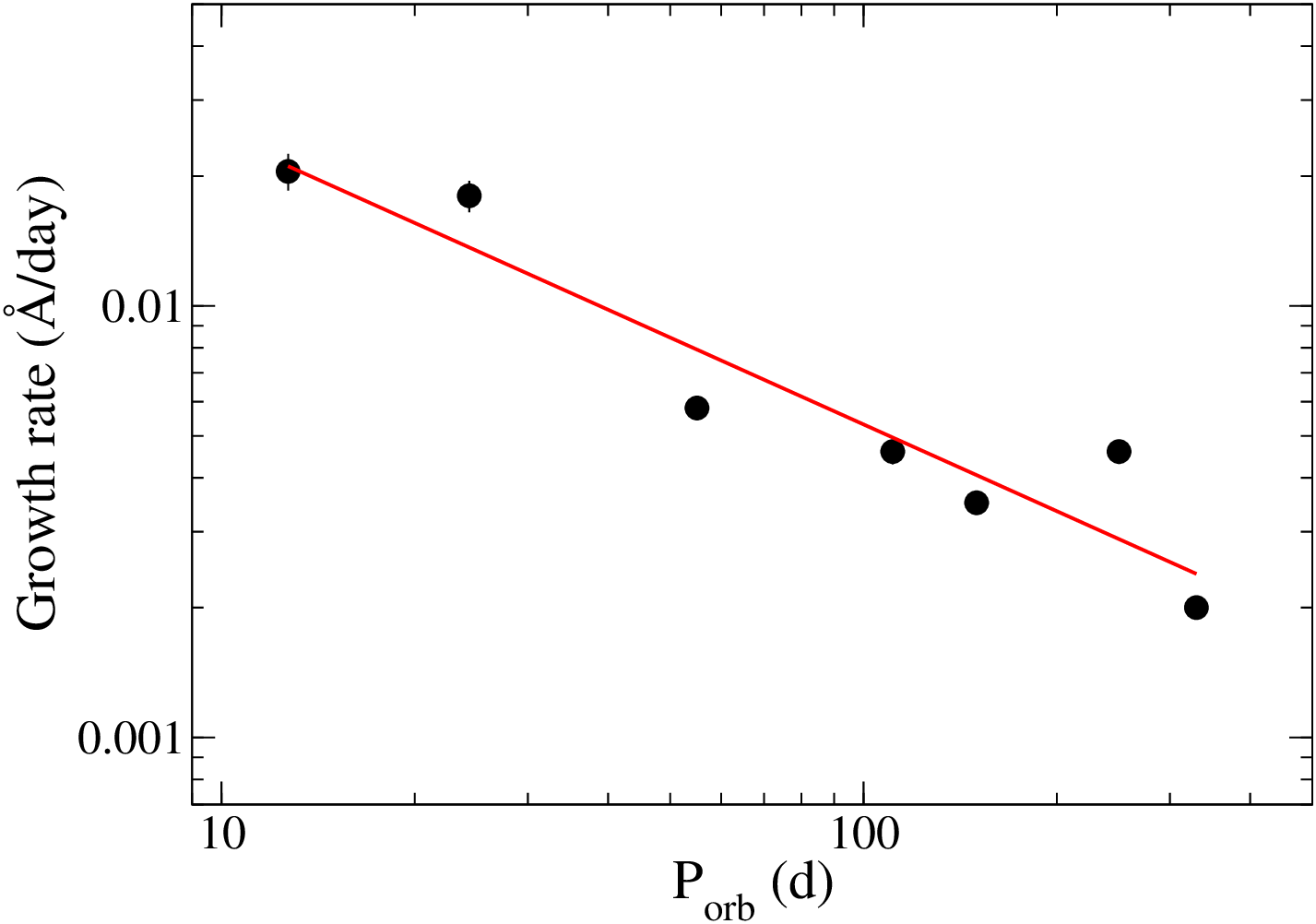}
\end{center}
\caption[]{Disk growth rates as a function of orbital period. The solid
line represents the best-fit to a power-law function.}
\label{growth}
\end{figure}

\subsection{Disk growth rates}
\label{diskgrowth}

We  computed the disk growth rates after the disk-loss episodes in 
SAX\,J2103.5+4545 (0.0205 \AA/d), 4U\,0115+63 (0.0180 \AA/d), 1A\,0535+262
(0.0058 \AA/d), RX\,J0440.9+4431 (0.0035 \AA/d), and X\,Per (0.0046 \AA/d)
and after the low states of GRO\,J2058+42 (0.0058 \AA/d) and
RX\,J0146.9+6121 (0.0020 \AA/d). The growth rates were calculated as the
slope of a linear fit of the \ew\ as a function of time between the minima
and maxima after the disk-loss event, representing states with no or very
weak disk and a full developed disk, respectively. We find  that the disk
in systems with short orbital periods grows faster than in systems where
the neutron star orbits further away.  The data fit  a power law,
$y=(0.12\pm0.01)\times x^{(-0.67\pm0.10)}$, well, as shown in Fig.~\ref{growth}. 

This result does not necessarily mean  that the mechanism that expels
matter from the photosphere of the Be star to create the disk, or even the
mass-loss rate from the star is different in different systems. The most
likely explanation is that, owing to truncation, the disk becomes denser more
rapidly in shorter orbital period systems, hence \ew\ changes faster.  As
the disk grows, the effect of the resonant torque from the neutron star
becomes stronger and the radial density distribution begins to  break at 
the truncation radius. Since the resonant torque prevents disk material
from drifting outwards, the disk density increases faster in systems where
the distance between the two components of the binary is shorter. This is
the same argument used by \citet{okazaki02}, which predicts that the disks in
BeXBs should be more dense than in classical, isolated Be stars. 

This correlation provides further evidence for the truncation of the disk
in BeXB systems. However, we advise caution given   the uncertainty in some of
the orbital periods of the systems in Fig.~\ref{growth} and the approximate
nature of the growth-rate calculation. The results  of this section should
be confirmed by future observations.

\begin{figure}
\begin{center}
\includegraphics[width=8cm]{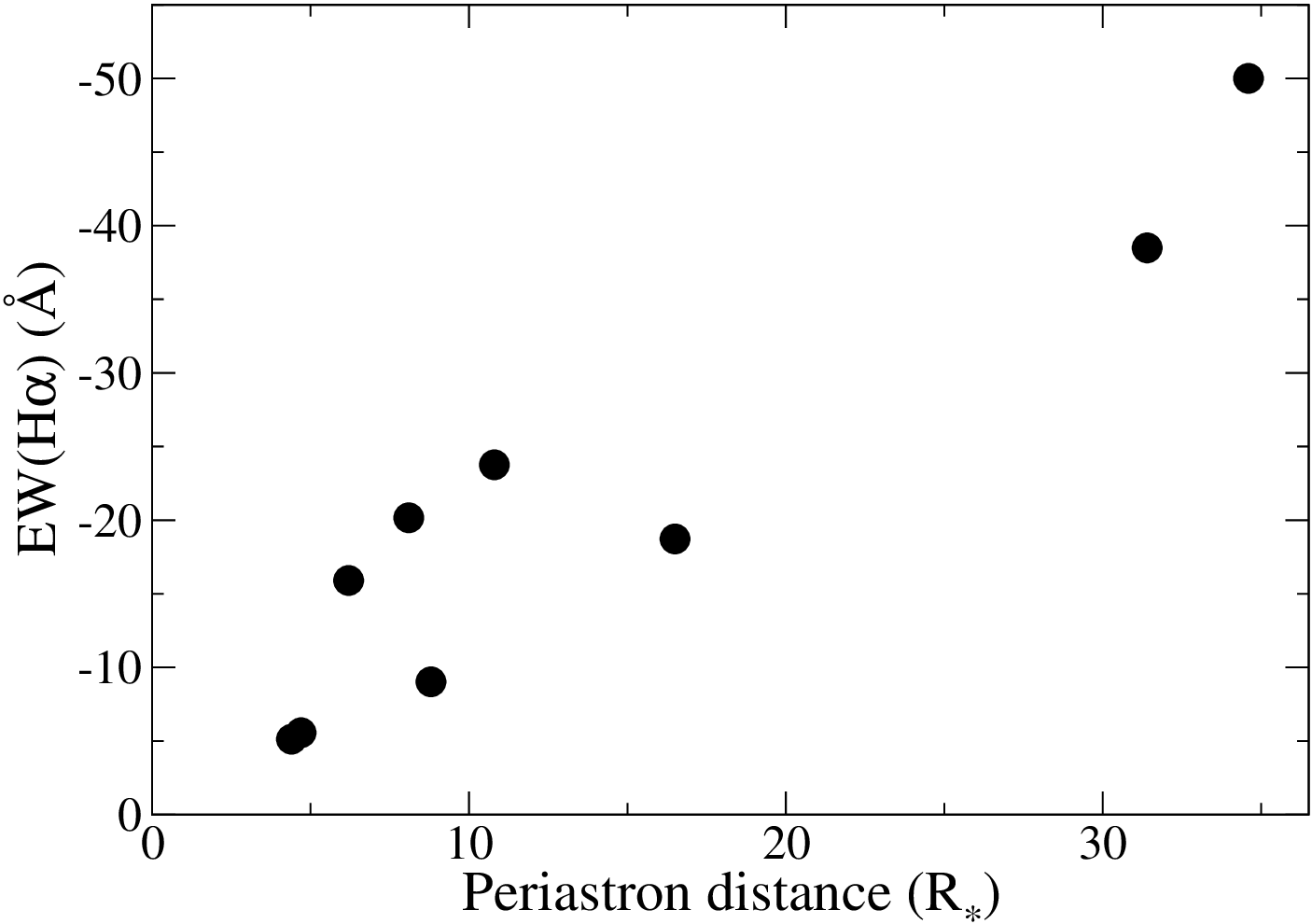}
\end{center}
\caption[]{Correlation between the periastron distance and the largest
value of \ew. }
\label{ew-per}
\end{figure}

\subsection{The equivalent width of the \ha\ line versus the orbital
separation}
\label{ewper}

Figure~\ref{ew-per} shows the correlation between the distance to
periastron and the maximum value of \ew. Periastron distances were computed
assuming the mass, radius, and orbital parameters given in
Table~\ref{targets}. Figure~\ref{ew-per} should be compared to the analogous
relationship between the orbital period and \ew\
\citep{reig97a,reig07a,antoniou09}.  The smaller the periastron distance,
the smaller the \ew.  The disk can grow until the torque exerted by the
neutron star begins to be strong, that is, until the disk radius reaches
the critical radius of the Roche lobe at periastron. The neutron star
removes angular momentum from the disk at each periastron passage, which
prevents the disk from expanding further. The closer the neutron star
passage, the stronger the tidal torque on the disk, hence the smaller
the disk and the lower the \ew. 

Quantifying these statements  depends on what one considers a large or
small disk. Assuming that \ew\ $\simmore -30$ \AA\ denotes a large disk,
then from Fig.~\ref{ew-per} we conclude that for a large disk to form requires
that the neutron star does not get closer than $\sim 20$ stellar radii
during periastron.

\begin{figure}
\begin{center}
\includegraphics[width=8cm]{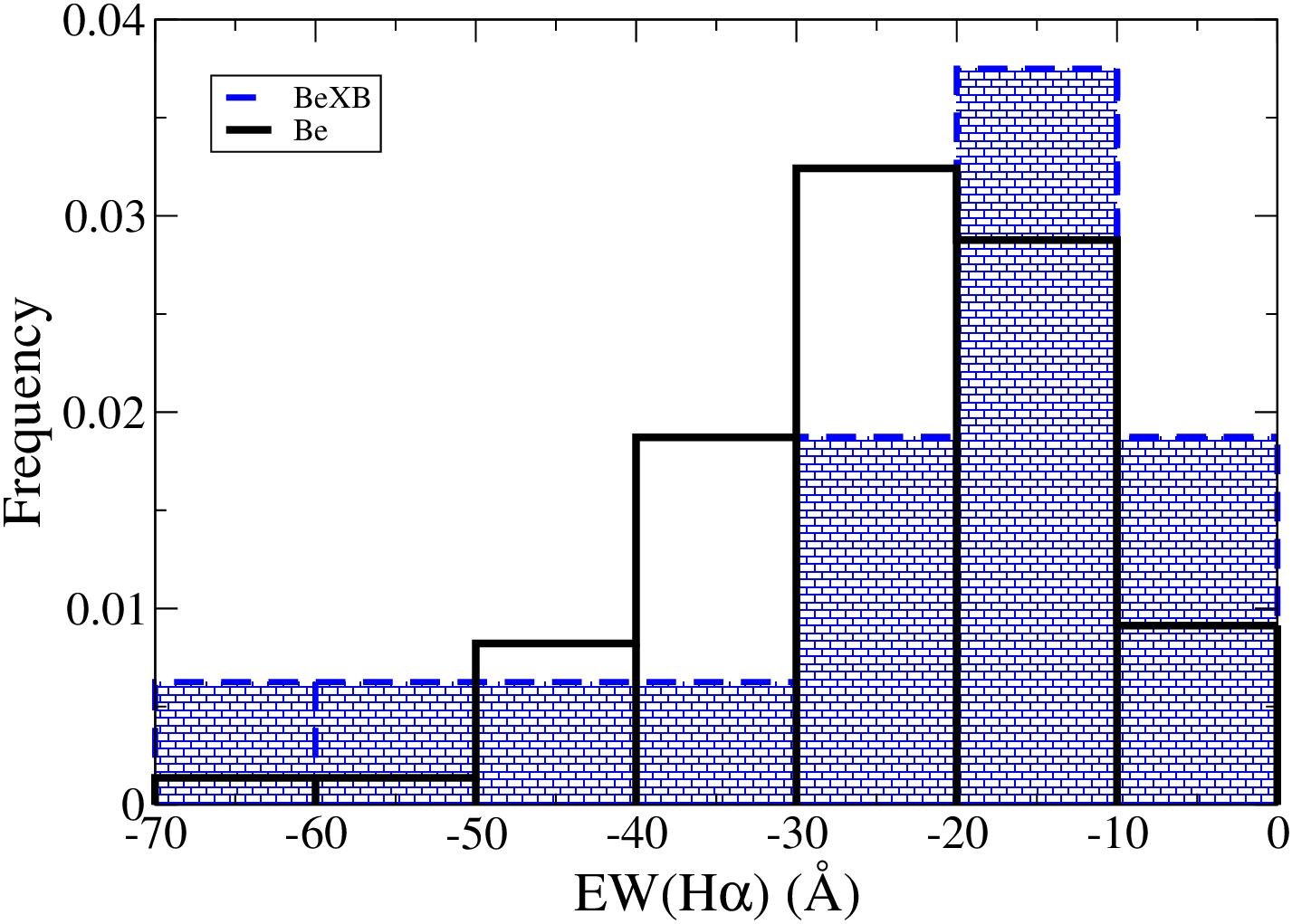}
\end{center}
\caption[]{Comparison of the \ew\ in classical Be stars (black line) and
BeXB (blue dashed hatched histograms). The data for the classical Be stars 
come from the IPHAS sample and comprises 219 stars with spectral types between 
O9 and B3. The data for BeXBs are from this work. }
\label{ew_hist}
\end{figure}

\subsection{The equivalent width of the \ha\ line in Be/X-ray binaries and
classical Be stars}
\label{hist}

Further evidence of the interaction between the neutron star and the disk
can be obtained by comparing the strength of the \ha\ emission in classical
Be stars and BeXB. Figure~\ref{ew_hist} shows a histogram of the equivalent
width in isolated Be stars and BeXBs. The data for BeXB systems comes from
the present work, and the values of \ew\ used to make Fig.~\ref{ew_hist}
were the largest measured so far, while the  \ew\ of  isolated Be stars was
taken from the INT Photometric \ha\ Survey (IPHAS) selected sample and
correspond to a single epoch \citep{gkouvelis16a,gkouvelis16b}. We used only
Be stars whose spectral type lies in the range O9--B3 to match those
observed in BeXB. The histogram shows the frequency, i.e. the number of
sources in the bin divided by the total number of sources.
Figure~\ref{ew_hist} was made using 219 Be stars and 16 BeXBs. Below \ew=--30
\AA, there is only one BeXB in each bin. This figure indicates that, on
average, isolated Be stars show distinctly larger \ew, hence larger disks. 

A Kolmogorov-Smirnov test of the two data sets gives a probability of 3.4\%
that they are drawn from the same distribution. We note however that, while
in BeXBs we have considered the largest value of \ew\ measured so far, for
classical Be stars we only have one measurement for each source.  Although
we cannot be sure that a larger \ew\ will not be measured in the BeXB
systems, the fact that the observations of BeXBs span  many years  makes it
unlikely, especially in cases where our observations cover several disk
dissipation-formation cycles, that future observations will  give 
significantly larger \ew. However, it is likely that many of the classical
Be stars will show larger \ew\ in the future because only one single epoch
measurement was used. Thus, the difference between the strength of the \ha\
line in classical Be stars and BeXBs is, in fact, more significant  than is
suggested by Fig.~\ref{ew_hist}.

\begin{table}
\caption{Statistics of the shape of the \ha\ line. The profiles observed
were classified into absorption (AB), single-peak (SP), fill-in (FI), 
double-peak (DP), multi-peak (MP), and P Cygni (PC).}
\label{shape}
\begin{center}
\begin{tabular}{lcccccc}
\hline  \hline
Source                  &AB     &SP     &FI     &DP       &MP   &PC   \\        
\hline
2S\,0114+65             &0      &31     &0      &0        &18    &8   \\
4U\,0115+63             &8      &0      &11     &87       &0     &0   \\
IGR\,J01363+6610        &0      &43     &0      &0        &0     &0   \\
RX\,J0146.9+6121        &0      &7      &0      &99       &0     &0   \\
IGR\,J01583+6713        &0      &30     &0      &0        &0     &0   \\
RX\,J0240.4+6112        &0      &0      &0      &44       &0     &0   \\
V\,0332+53              &0      &42     &0      &11       &0     &0   \\
X\,Per                  &4      &45     &0      &27       &0     &0   \\
RX\,J0440.9+4431        &0      &0      &0      &65       &0     &0   \\
1A\,0535+262            &2      &26     &0      &59       &0     &0   \\
IGR\,J06074+2205        &7      &0      &0      &29       &0     &0   \\
AX\,J1845.0-0433$^\dag$ &0      &0      &0      &0        &29    &(29)\\ 
4U\,1907+09             &0      &3      &0      &0        &43    &0   \\
XTE\,J1946+274          &0      &26     &0      &0        &0     &0   \\
KS\,1947+300            &0      &50     &0      &1        &0     &0   \\
GRO\,J2058+42           &0      &0      &1      &57       &0     &0   \\
SAX\,J2103.5+4545       &47     &0      &0      &27       &0     &0   \\
IGR\,J21343+4738        &12     &0      &0      &14       &0     &0   \\
4U\,2206+54             &0      &0      &0      &66       &0     &0   \\
SAX\,J2239.3+6116       &0      &0      &0      &26       &0     &0   \\
\hline
Total (SGXB)            &0      &34     &0      &0        &90    &36  \\
Total (BeXB)            &80     &401    &12     &612      &0     &0   \\
\hline
\hline
\multicolumn{7}{l}{$\dag$: These profiles show both a multi-peak and a P
Cygni shape.} \\
\end{tabular}
\end{center}
\end{table}

\subsection{Profile types. V/R peak height variability}
\label{profvar}

While the strength of the \ha\ line provides information about the size of
the emitting region, its morphology indicates whether the distribution of
gas particles in the disk is homogeneous. Symmetric lines are believed to
be generated in axisymmetric disks. Asymmetric lines are due to a
global density perturbation that revolves inside the disk \citep[see
e.g.][and references therein]{hummel95}. Transitions from one type of line
to the other are common among classical Be stars.

\begin{figure}[t]
\includegraphics[width=8cm]{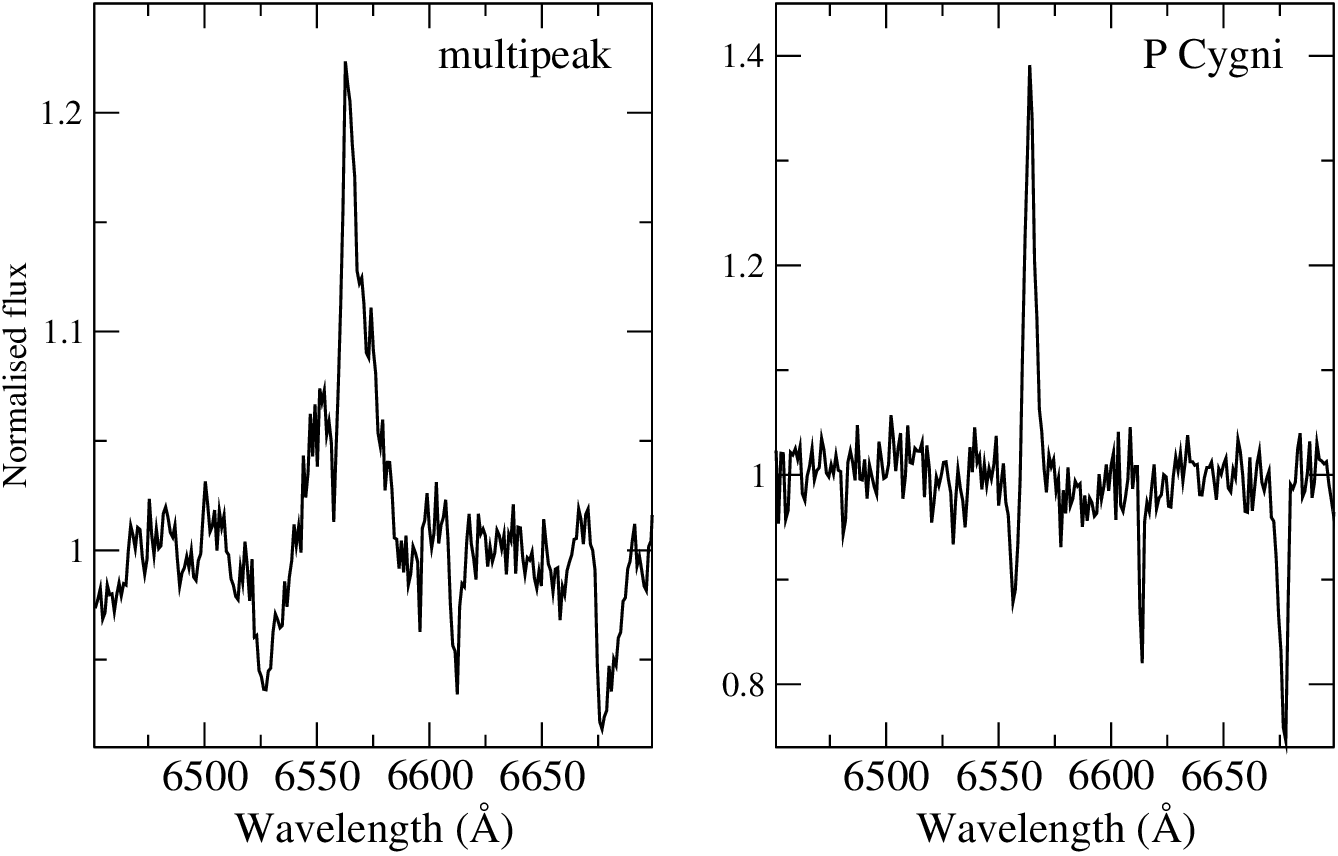}
\caption[]{Representative \ha\ profiles in SGXBs: {\em right}: multipeak
profile observed in AX\,J1845.0-0433; {\em left}: P Cygni profile observed
in 2S\,0114+65.}
\label{profile-sgxb}
\end{figure}

\begin{figure}[t]
\includegraphics[width=8cm]{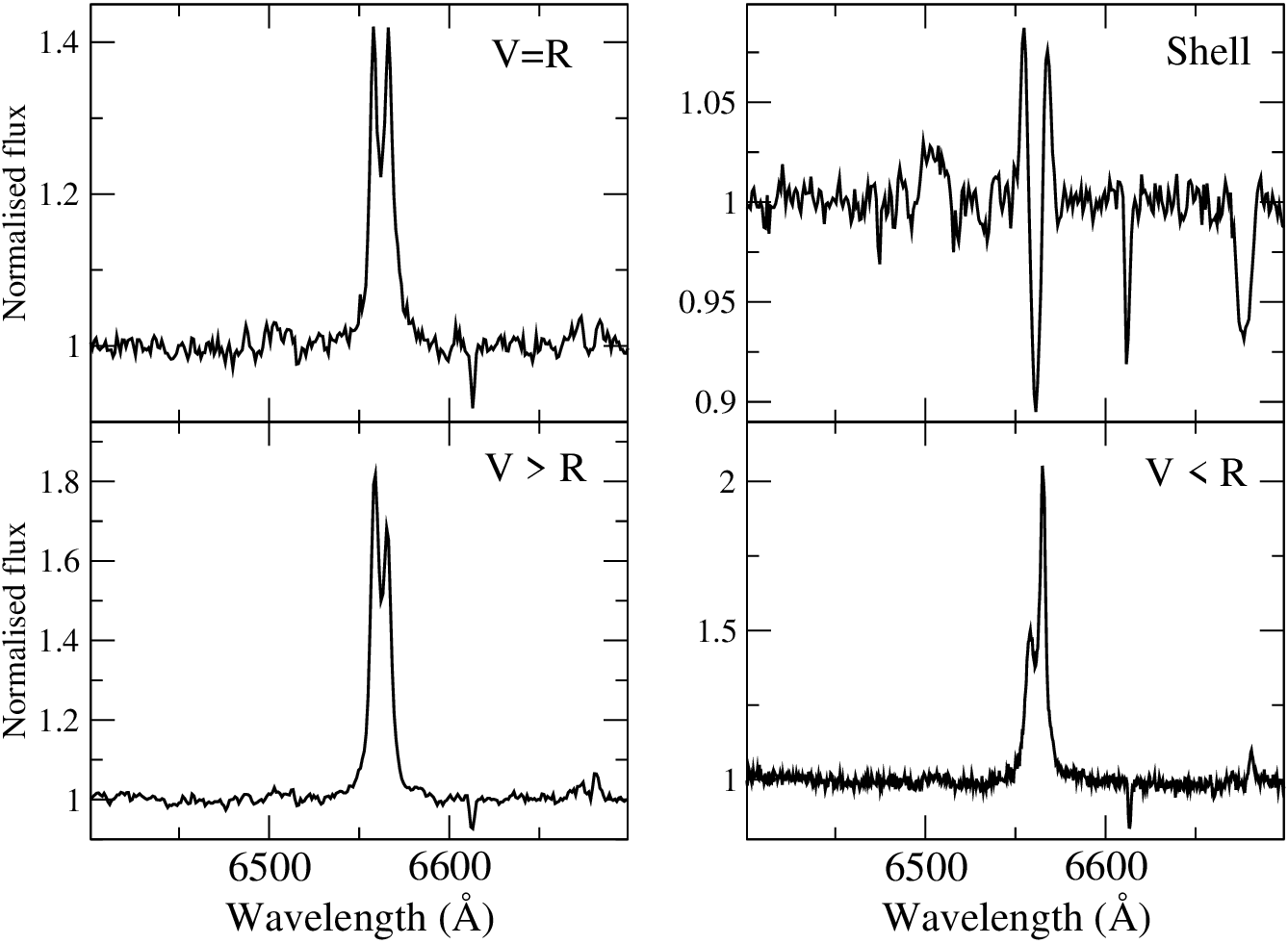}
\caption[]{Representative double-peak \ha\ profiles in BeXB (from
RX\,J0440.9+4431).}
\label{profile-bexb}
\end{figure}

Table~\ref{shape} summarises the number of times that the different
profiles have been observed: absorption (AB), fill-in (FI), single-peak
(SP), double-peak (DP), multi-peak (MP), and P Cygni (PC). Fill-in profiles
are those cases in which the line has been filled with emission by an
amount that approximately covers the photospheric absorption, resulting in
a flat spectrum at the position of the line, hence EW(H$\alpha$)$\approx$
0. The DP type also includes  the so-called shell profiles. These are
double-peak profiles in which the central absorption that separates the two
peaks reaches or goes below the continuum. By multi-peak, we mean a profile
that shows a complex structure, typically a main peak whose flanks  are
affected by smaller peaks  or weak absorption features. Finally, P Cygni
profiles are emission lines with a blue-shifted absorption component, which
is produced by material moving away from the star (i.e. stellar wind)
towards the observer's line of sight. Representative \ha\ line profiles for
SGXBs and BeXBs are given in Fig~\ref{profile-sgxb} and
Fig.~\ref{profile-bexb}, respectively.

Although the shape of the line greatly depends on the spectral resolution,
a statistical analysis of the various types of profiles may still be useful
to study the connection of the line profile with other properties of the
system and gain some insight into the mass transfer process.  Taking
into account the resolution of our spectra (typically $R\simless 2500$), an
inspection of Table~\ref{shape} reveals the following general results:
{\em i)} which systems went through disk-loss episodes (those showing
absorption profiles), {\em ii)} that MP and PC types are only seen in
SGXBs\footnote{Although some BeXBs may show multi-peaked structures,
these peaks normally appear on the top of the line and are observed at very
high resolution \citep{moritani13}. The multi-peaked profiles that we
observe in the SGXBs of our list show peaks or absorption features not only
on the top but also on the flanks of the line.}, {\em iii)} that a
double-peak profile is the dominant shape in most BeXBs,  especially at
low and intermediate \ew, and {\em iv)} together with Table~\ref{minmax},
that large \ew\  in BeXBs tends to be associated with single peak
profiles. Items {\em iii)} and {\em iv)} are a natural consequence of
emission lines whose widths essentially have  a kinematic origin
\citep{hanuschik89}.

The most prominent spectroscopic evidence of disk activity is the long-term
V/R variability, which refers to the variation in the relative intensity of
the blue (V) and red (R) peaks in the split profile of the line. In
long-lasting disks, it is possible to observe more than one cycle $V > R
\longrightarrow V \approx R \longrightarrow V < R$. However, owing to the
very nature of the source of variability, namely, the precession of a
density perturbation in the disk, we do not expect the process to be
precisely periodic with a high degree of coherence. The duration of the V/R
cycle presumably depends on disk parameters such as viscosity, density,
and size, which are not only  poorly known but also change as the disk grows. In
addition, observational gaps in the data introduce extra complications.
Therefore software packages designed to search for periodicities in data do
not generally produce a significant result. Nevertheless, by isolating
periods where the V/R cycle appears to be coherent and using algorithms
that can handle gaps (e.g. Lomb-Scargle), it is possible to obtain the V/R
quasiperiods. Otherwise, we just estimated the quasiperiod of the data by eye.

\begin{figure*}
\begin{center}
\includegraphics[width=16cm]{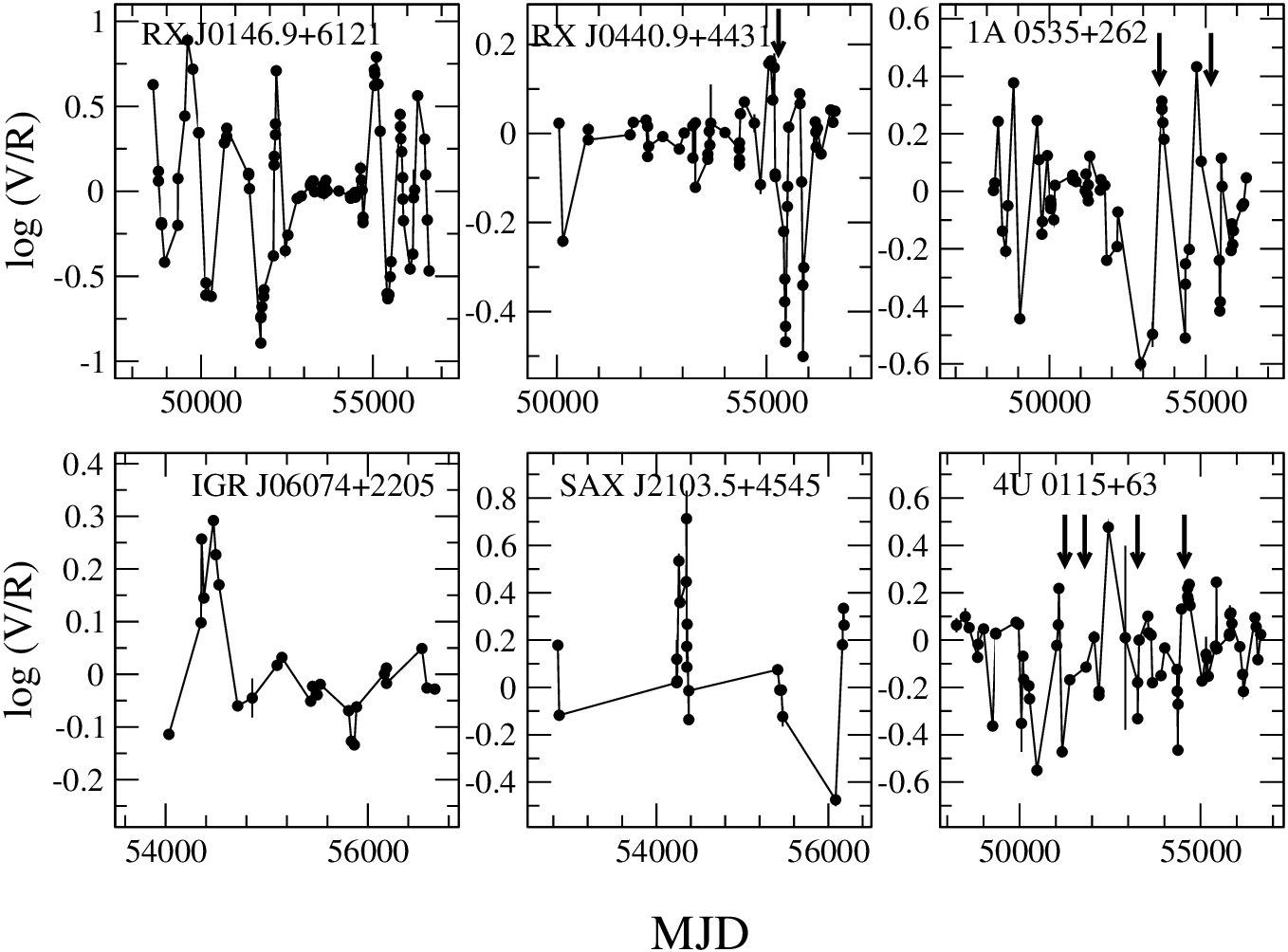}
\end{center}
\caption[]{Variability of the \ha\ line profile quantified as the $\log(V/R)$, 
where $V$ and $R$ are the intensity of the left and right peaks. Arrows
indicate the occurrence of large X-ray outbursts.}
\label{vr}
\end{figure*}

All but three sources of our BeXB sample\footnote{A split profile is
observed in IGR\,J01583+6713 only in the higher resolution spectra. In
these cases, the peak separation is small, $\simless$ 100 km s$^{-1}$ and $V
\approx R$.} (IGR\,J01363+6610, XTE\,J1946+274, and KS\,1947+300) showed double-peaked profiles at
some point, whereas SGXBs show broad (FWHM$\sim$
300--900 km s$^{-1}$) multi-peak lines.  4U\,2206+54 consistently shows a
slightly red dominated profile with an almost constant ratio of
$V/R=0.8\pm0.1$. In RX\,J0240.4+6112, only four spectra have $V > R$, while
the remaining 40 have $V<R$, also with an average ratio of  $V/R=0.8\pm0.1$.
Fig~\ref{vr} shows the evolution of  $\log(V/R)$ for the most variable
systems. Of particular interest are  systems 1A\,0535+262,
RX\,J0440.9+4431, 4U\,0115+63, and RX\,J0146.9+6121. 

RX\,J0146.9+6121 was in a bright optical state between March 1993 and
October 2001 (MJD 49000--52500). \ew\ remained below --8 \AA\ reaching a
historical maximum of --20 \AA\ in September 1993 (Fig.~\ref{diskew}). Then
\ew\ began to decrease until July-August 2005, when the lowest value of
--3.7 \AA\ was recorded. During the faint state (MJD 53000--54800), the line
showed a symmetric double-peaked profile with $V=R$ (Fig.~\ref{vr}). When
the \ew\ began to increase again, the asymmetries returned, starting with
$V<R$. Before the disk minimum, we computed a V/R quasi-period of
$\sim$1300 d (MJD 49000--52500) while, after the minimum, the quasi-period
was half $\sim$650 d  (MJD 54700--56600). The shorter period after the
faint state can be attributed to a smaller disk: initially the disk was
large, as indicated by the larger values of \ew. Slowly and progressively
the disk decreased in size to a minimum. At this time, the disk is so small
that it cannot support the perturbation, which vanishes. Without a perturbed
disk, the V/R ratio becomes close to 1 ($V \approx R$), i.e. a symmetric
profile. As soon as the disk
recovers a certain size, the perturbation appears again, but because the disk
is smaller than previously, the perturbation revolves faster, resulting in
a smaller period.

The case of RX\,J0440.9+4431, 1A\,0535+262, and 4U\,0115+63 is similar. 
After a disk-loss episode (see Fig.~\ref{diskew}), the disk reformed
slowly, with the \ew\ increasing progressively. Once the disk attained a
large size, V/R variability set in.  We measured a quasiperiod of $\sim$525
d in RX\,J0440.9+4431 and $\sim$1100 d in 1A\,0535+262. In 4U\,0115+63 the
disk forms and dissipates on timescales of 3--5 years
\citep[Fig.~\ref{diskew}, see also][]{reig07b} and the V/R quasiperiod
(1000--1500 d) is bound by these fast changes. Eventually, the growth of
the disk in these three systems  led to a major (type II) outburst.  An
interesting result is the fact that the asymmetric profiles are observed
{\em before} the onset of the X-ray outburst, supporting the models that
invoke highly distorted disks to explain Type II accretion events
\citep{okazaki13, martin14}

Although the disk in IGR\,J06074+2205 and SAX\,J2103.5+4545 does not reach
a stable configuration for a long enough time to measure a periodicity, the
two systems display fast V/R variability with the V/R ratio reverting from
V/R $>$ 1 to V/R $<$ 1 on timescales of $\simless 1$ year.

\begin{figure*}
\begin{center}
\includegraphics[width=16cm]{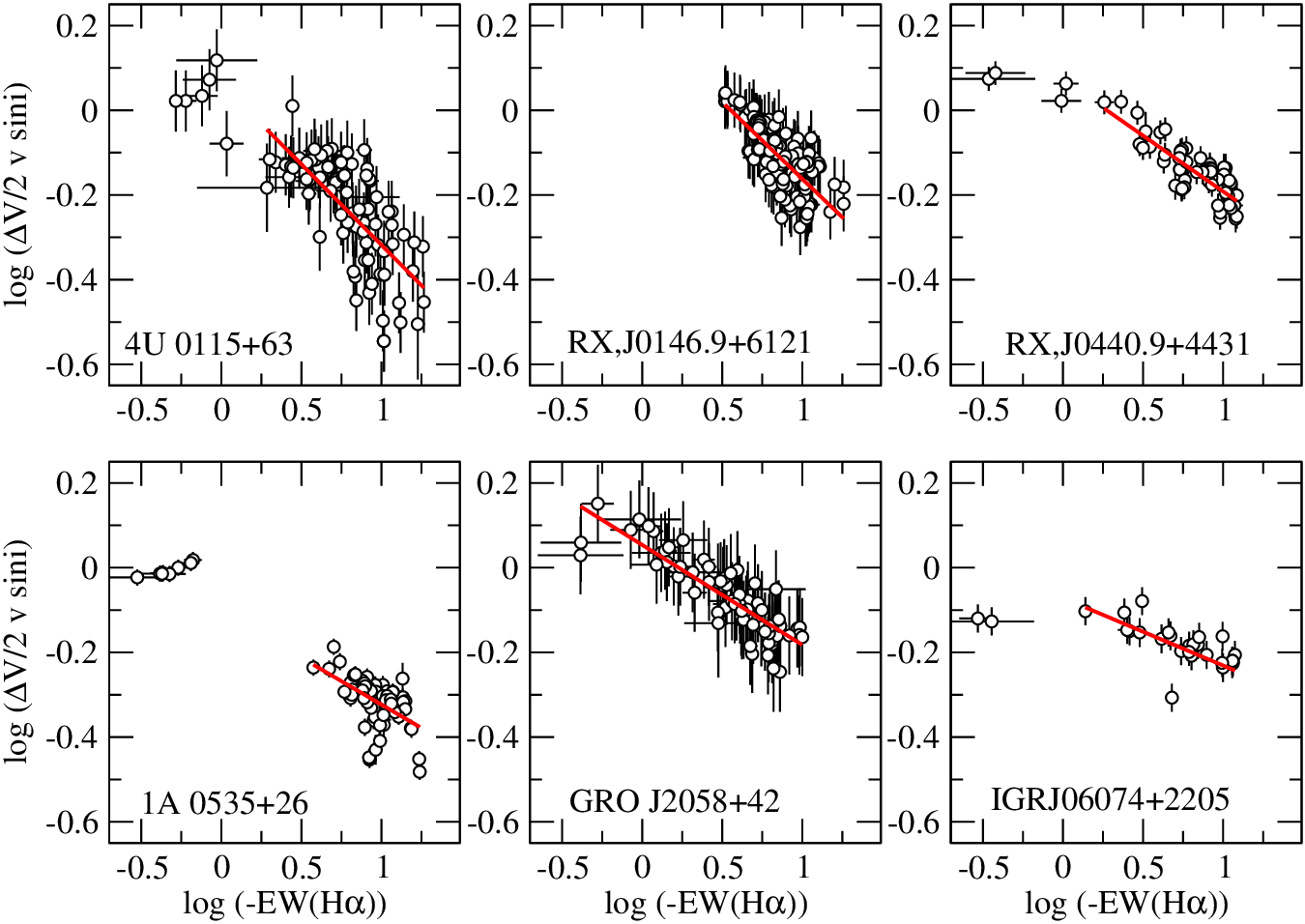}
\end{center}
\caption[]{Correlation between the peak separation and the \ha\ equivalent
width for six BeXBs. The line represents the best fit to a linear regression
(see Table~\ref{rotlaw}). The scale of the axes was the same in all 
plots to facilitate comparison.}
\label{ewdelta}
\end{figure*}

\subsection{Disk velocity law}
\label{diskvelaw}

A long standing issue since the discovery of the first Be stars has been
the nature of the velocity law that is followed by the particles in the disk.  The 
correlations between the various spectral parameters of the \ha\ line (FWHM, EW, peak
separation in double-peaked profiles) and with the stellar
rotational velocity that have been observed in many Be stars are
interpreted as evidence for rotationally dominated circumstellar disks
\citep{dachs86b,hanuschik89,dachs92}.  The velocity profile
in a circumstellar disk can be expressed as

\begin{equation}
\label{velaw}
v_{\rm rot}=v_*\left(\frac{R_*}{r_{\rm d}}\right)^{j}
,\end{equation}

\noindent where $v_*$ and $R_*$ are the star rotational velocity and radius
and $j$ is the disk rotational velocity law parameter, which may adopt
three values \citep[see e.g.][]{dachs92}: {\em i)}  $j=1$ represents the
case of conservation of angular momentum, where the circumstellar gas has
the same angular momentum per unit mass as that at the stellar surface,
{\em ii)} $j=-1$ corresponds to rigid rotation, which might occur in the
innermost regions of the disk, and {\em iii)} $j=0.5$ for a Keplerian
disk.

Nowadays, there is general consensus, based on the evidence gathered over
many years,  that the circumstellar disks in classical Be stars are
governed by viscosity and follow a Keplerian law
\citep{meilland07,rivinius13a}. A way to estimate the index $j$, and hence
determine the type of dependence of the rotational velocity with radius, is
through the relationship between the peak separation, $\Delta V$, and the
equivalent width, \ew.  \citet{hanuschik88} derived the law

\begin{equation}
\label{jindex}
\log \left(\frac{\Delta V}{2 v_* \sin i}\right)= -a
\log\left(-EW(H\alpha)\right)+b
,\end{equation}

\noindent where $i$ is the inclination angle.  The equivalent width is
expected to be  proportional  to the visible disk area, i.e. to the disk
radius squared \citep{tycner05,grundstrom06}. In this case,  $a=j/2$ and b
is related to the electron density in the disk \citep{hanuschik88}. From a
sample of 93 isolated Be stars of all spectral types \citet{hanuschik89}
found $a=0.32$ and $b=-0.20$. We repeated this calculation using  the IPHAS
list of isolated Be stars \citep{gkouvelis16b}. Of the 230 Be stars with
spectral types in the range O9--B3, only 30 displayed double peaked
profiles. A fit to this data gave $a=0.24\pm0.11$ and $b=-0.15\pm0.12$. For
the IPHAS sample of classical Be stars, $j\approx 0.5$, as expected for a
Keplerian disk. 

The question that we wish to address here is whether the disks of Be stars
in X-ray binaries also follow a Keplerian law or, instead, the presence of a
neutron star somehow alters the expected behaviour. Figure~\ref{ewdelta}
shows the correlation between the peak separation and the equivalent width
of the \ha\ line for six BeXBs. The result of the fit and the value of the
$j$ index are given in Table~\ref{rotlaw}. The correlation is expected to
hold for well developed disks, hence data points with $\log(EW(H\alpha))<0$
were ignored in the fit. None of the sources deviate by more than 2$\sigma$
from the value of $j=0.5$. Hence we conclude that the circumstellar disks
in BeXBs are also Keplerian. The only exception is  IGR\,J06074+2205, which
might be explained by the fast timescales for disk formation and
dissipation. Although the data span  almost eight years, this source
never reaches a stable state with a well developed disk for more than a
year (see Fig.~\ref{diskew}). Thus most of the data may correspond to
phases where the disk is forming or dissipating, where a Keplerian supported
disk may not be at work.

\begin{table}
\caption{Linear regression between the peak separation and the equivalent width
of the \ha\ line.  All correlations are significant at $>99.999$\%.}
\label{rotlaw}
\begin{center}
\begin{tabular}{lccc}
\hline  \hline
Source name        &$j$                 &Intercept              &Corr. coeff.   \\
\hline
4U\,0115+634       &0.56$\pm$0.08       &0.003$\pm$0.032        &--0.69\\
RX\,J0146.9+6121   &0.70$\pm$0.10       &0.20$\pm$0.04          &--0.77\\
RX\,J0440.9+4431   &0.54$\pm$0.03       &0.08$\pm$0.01          &--0.86\\
1A\,0535+262       &0.43$\pm$0.04       &--0.10$\pm$0.02        &--0.71 \\
IGR\,J06074+2205   &0.28$\pm$0.06       &--0.07$\pm$0.02        &--0.82 \\
GRO\,J2058+42      &0.52$\pm$0.08       &0.07$\pm$0.03          &--0.86 \\
\hline
\end{tabular}
\end{center}
\end{table}

Although the slope of the correlation, i.e, the $j$ index,  is similar in
classical Be stars and BeXBs, the intercept is clearly different.  Figure
\ref{densedisk} again shows  the relationship between the \ew\ and the peak
separation of our BeXB, but this time all the data are plotted together. The
solid line represents the average behaviour of the 30 IPHAS classical Be stars
with spectral types in the range O9--B3 and the dotted line corresponds to
93 classical Be stars of all spectral types \citep{hanuschik89}. The
best-fit to the BeXB data points is given by the dashed red line.

As can be seen, BeXBs lie above the mean behaviour of classical Be stars,
i.e. they appear shifted vertically with respect the Be star line fit. 
According to \citet{hanuschik88}, the intercept $b$ is related to disk
density. Thus this shift implies that the circumstellar disks in BeXBs are
denser than disks in isolated Be stars.  We find a difference of $
\log(\Delta V/2 v \sin i)_{\rm BeXB} - \log(\Delta V/2 v \sin i)_{\rm Be}
\approx 0.15-0.2$, implying that  the disks of BeXB systems are about 1.5
times denser  than those of classical Be stars. This result agrees with
\citet{zamanov01} who found that on average the disks of BeXBs are about
twice as dense as those of classical Be stars, a result that was attributed
to disk truncation.

\subsection{Circumstellar disk radius}
\label{diskradius}

According to \citet{huang72}, the separation of the emission line peaks can
be interpreted as the outer radius of the emission line forming
region\footnote{Because the disk is not expected to terminate abruptly, the
radius obtained from the optically thick region where the \ha\ line arises
represents a lower limit to the true disk radius.}. We can estimate the
outer radius from Eq.~(\ref{velaw}) and the relationship between the peak
separation and the velocity at which gas rotates in the disk $\Delta V= 2
v_{\rm rot} \sin i$, where $i$ is the inclination angle. Isolating $r$ from
equation (\ref{velaw}), the radius of the \ha\ emitting disk is given by

\begin{equation}
\label{rad}
\frac{r_{\rm d}}{R_*}= \left(\frac{2 v_* \sin i}{\Delta V}\right)^{1/j} 
.\end{equation}

\begin{figure}
\begin{center}
\includegraphics[width=8.5cm]{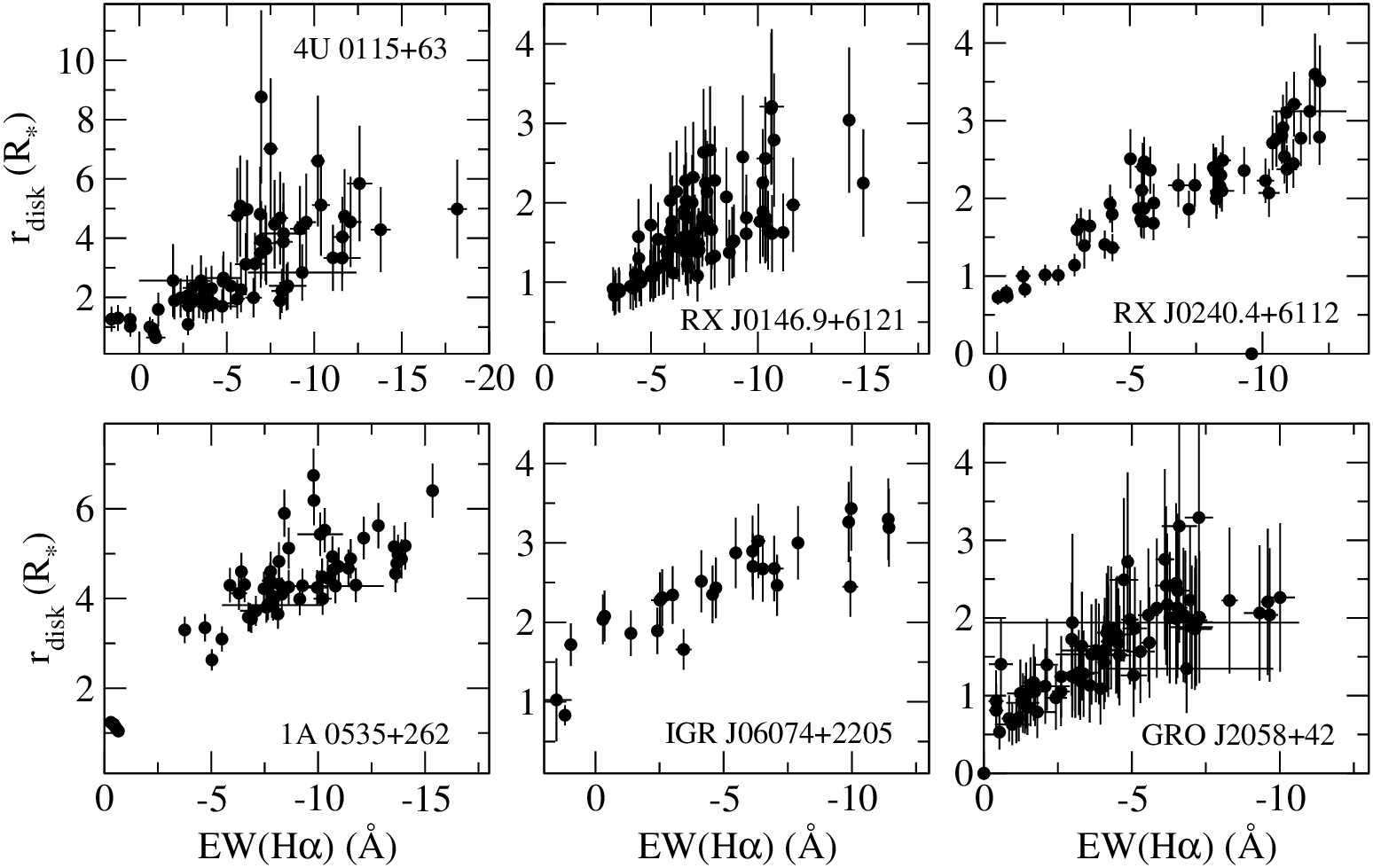}
\end{center}
\caption[]{Relationship between \ha\ equivalent width and disk radius.}
\label{rad-ew}
\end{figure}

We used Eq.~(\ref{rad}) and $j=1/2$ to estimate the radius of the
line-emitting region for those systems with known values of the star
projected rotational velocity and studied the relationship with \ew. A
positive correlation between \ew\ and the disk radius is found (see
Fig.~\ref{rad-ew}). This relationship confirms the well-known result that
\ew\ gives a measure of the size of the disk
\citep{quirrenbach97,grundstrom06}.

\begin{figure}
\begin{center}
\includegraphics[width=8.5cm]{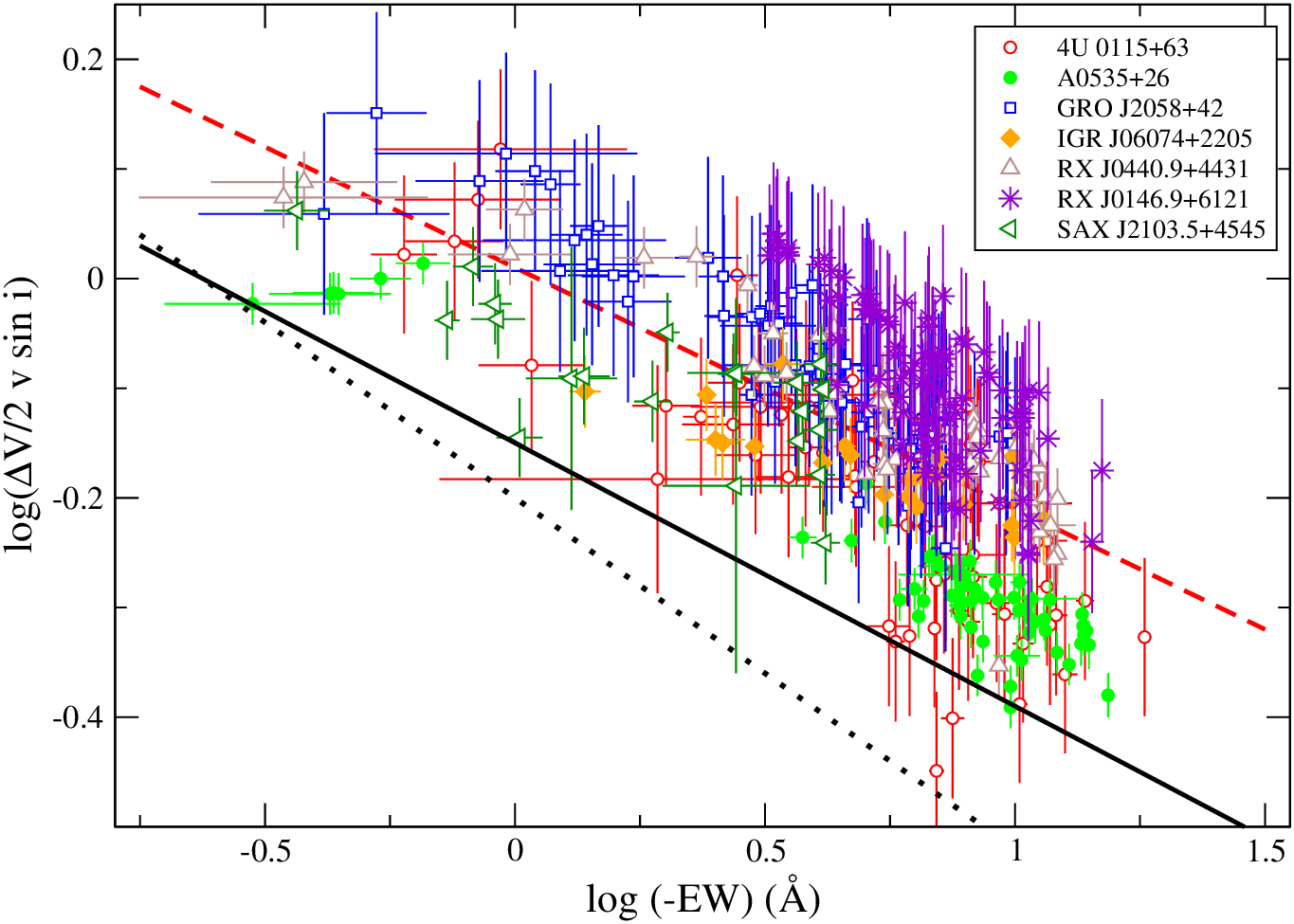}
\end{center}
\caption[]{Relationship between the peak separation and the \ha\ equivalent
width. The solid and dashed line represents the average behaviour of
30 classical Be stars with spectral type in the range O9--B3 
\citep{gkouvelis16b} and 93 Be stars of all spectral types
\citep{hanuschik89}, respectively. BeXBs lie systematically above the Be
stars' lines.}
\label{densedisk}
\end{figure}

We note that the anticorrelation between $\Delta V$ and \ew,
see Figs.~\ref{ewdelta} and \ref{densedisk}, rules out a velocity law with
$j<0$. Moreover, this anticorrelation means that, as the disk
radius or, equivalently, \ew, increases, the peak separation decreases until
the two peaks merge and a single peak is observed. Indeed, systems that
display large amplitude changes in \ew\ show single-peak profiles when \ew\
is large and double-peak profiles when \ew\ is small. The transition from
double- to single-peak profiles occurs when \ew\ increases above $\sim -15$ \AA,
although this value is expected to depend on the spectral resolution.

From Fig.~\ref{rad-ew}, we see that disk radii in BeXBs are typically $R_{\rm
disk}\simless 6 R_*$. This result can be compared with the average radii of
the \ha-emitting regions in classical Be stars: $<R_{\rm disk}>=14\,R_*$
\citep{hummel95},  $<R_{\rm disk}>=22\,R_*$ \citep{dachs92},  $<R_{\rm
disk}>=17\,R_*$ \citep{hanuschik86}, and $<R_{\rm disk}>=20\,R_*$
\citep{hanuschik88}. BeXBs have smaller disks than classical Be stars, in
agreement with the results presented in Sect.~\ref{hist}.


\subsection{Circumstellar disk and X-ray variability}

Because the disk constitutes the main reservoir of matter available for
accretion, a correlation between X-rays and optical emission should be
expected. X-ray outbursts in BeXB are commonly classified into a simple
scheme of  normal (Type I) and  giant (Type II) outbursts:

\begin{itemize}

\item Type I outbursts are regular and (quasi-)periodic outbursts, normally
peaking at or close to periastron passage of the neutron star. They are
short-lived, i.e. tend to cover a relatively small fraction of the orbital
period (typically 0.2--0.3 $P_{\rm orb}$). The X-ray flux increases by up
to two orders of magnitude with respect to the pre-outburst state, reaching
peak luminosities $L_X \simless 10^{37}$ erg s$^{-1}$.  The number of
detected Type I outbursts varies from source to source but typically the
source remains active for 3--8 periastron passages. A remarkable case is
EXO 2030+375, which shows type I outbursts almost permanently
\citep{wilson02}.

\item Type II outbursts represent major increases of the X-ray flux,
$10^3-10^4$ times that at quiescence. They may reach the Eddington
luminosity for a neutron star ($L_X \sim 10^{38}$ erg s$^{-1}$) and become
the brightest objects of the X-ray sky. They do not show any preferred
orbital phase and last for a large fraction of an orbital period or even
for several orbital periods. 

\end{itemize}

The standard picture is that Type I X-ray outbursts are triggered by
the mass transfer from a tidally truncated  Be
star's disk \citep{negueruela01a,okaneg01,okazaki01} at or near periastron,
while Type II outbursts are associated with unstable disk configurations
such as warped disks.
A recent work by \citet{okazaki13} shows that the standard disk model
cannot explain the large and rapid X-ray flux variations and the short
duration of normal Type I outbursts  because the accretion timescales are
expected to extend beyond a few orbital periods. \citet{okazaki13} propose
that normal outbursts can be explained by radiatively inefficient accretion
flows, while giant outbursts are due to Bondi-Hoyle-Lyttleton  accretion of
material transferred from the outermost part of a Be disk that is misaligned with
the binary orbital plane.  \citet{martin14} find that, in addition to
being warped and highly misaligned, the disk has to be eccentric to explain
giant outbursts.

In this section, we concentrate on the changes experienced by the disk {\em
after} a major X-ray outburst. Because of the good correlation between disk
radius and \ew\ and because \ew\ is more easily measured, we used the
equivalent width rather than the disk radius to investigate the correlation
between the optical and X-ray variability. Moreover, the radius can be
computed only when a double peak is observed, hence the use of \ew\ provide
more data points and more sources where the relationship can be studied.

\begin{figure*}
\begin{center}
\begin{tabular}{cccc}
\includegraphics[width=7cm]{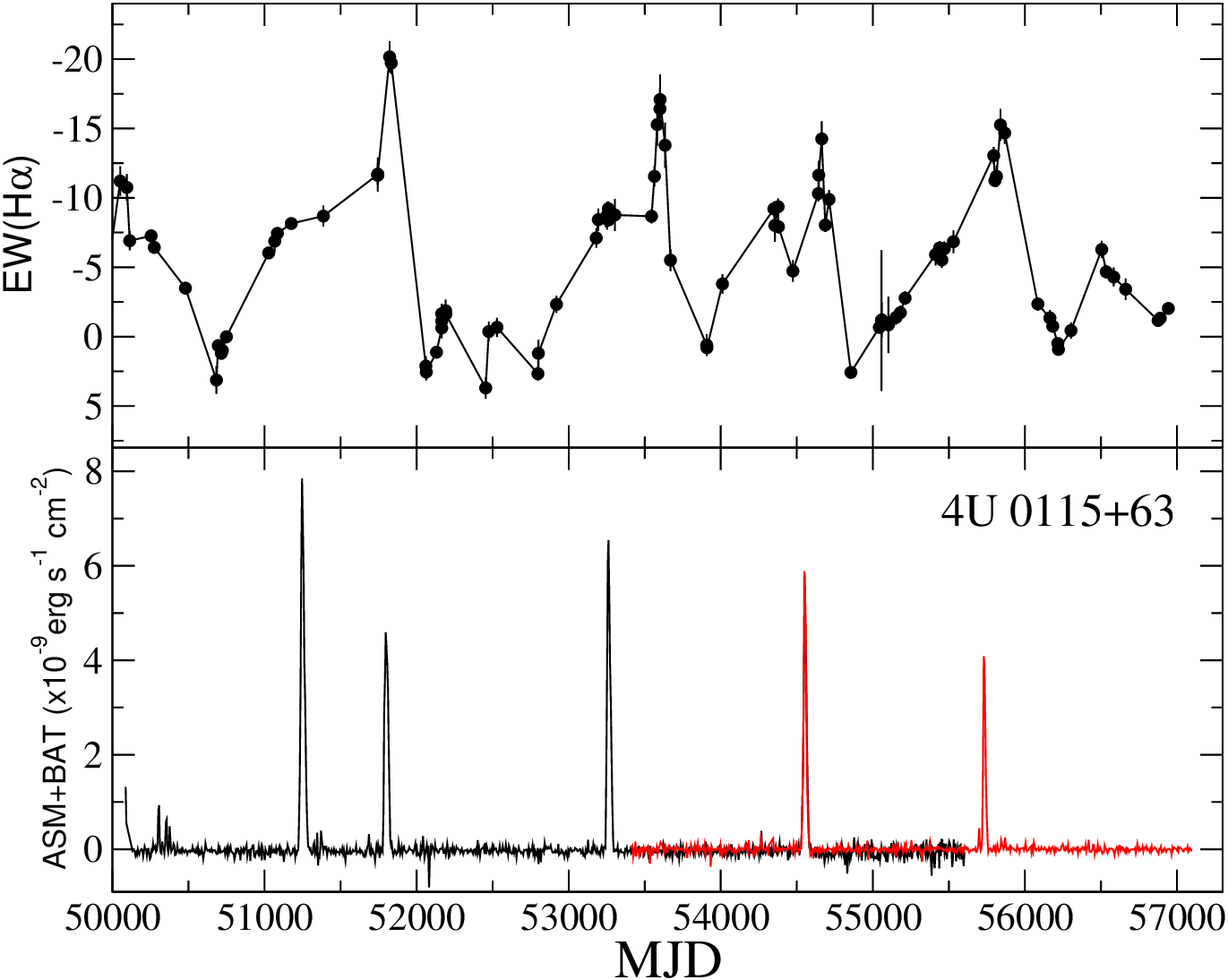} &
\includegraphics[width=7cm]{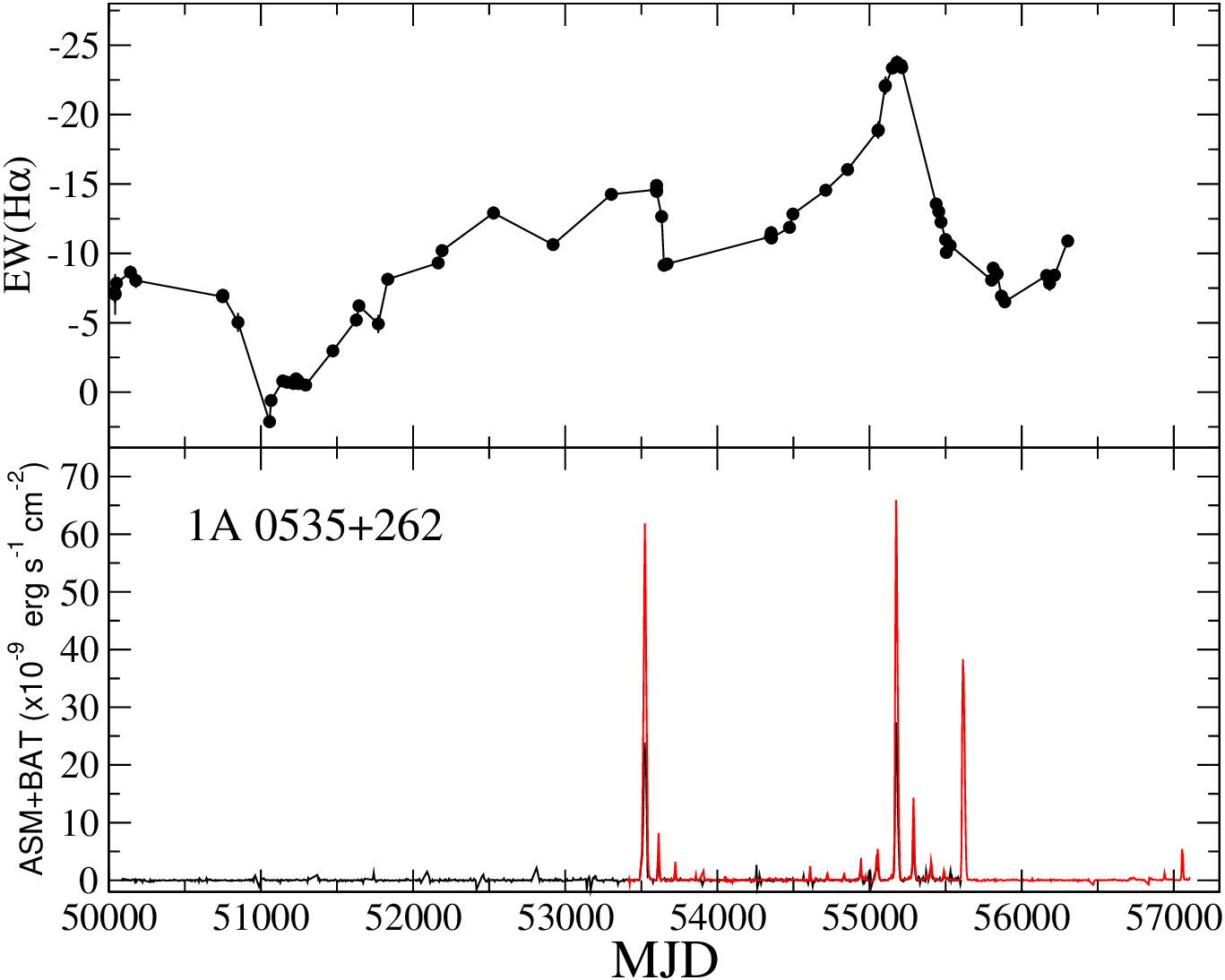}  \\
\includegraphics[width=7cm]{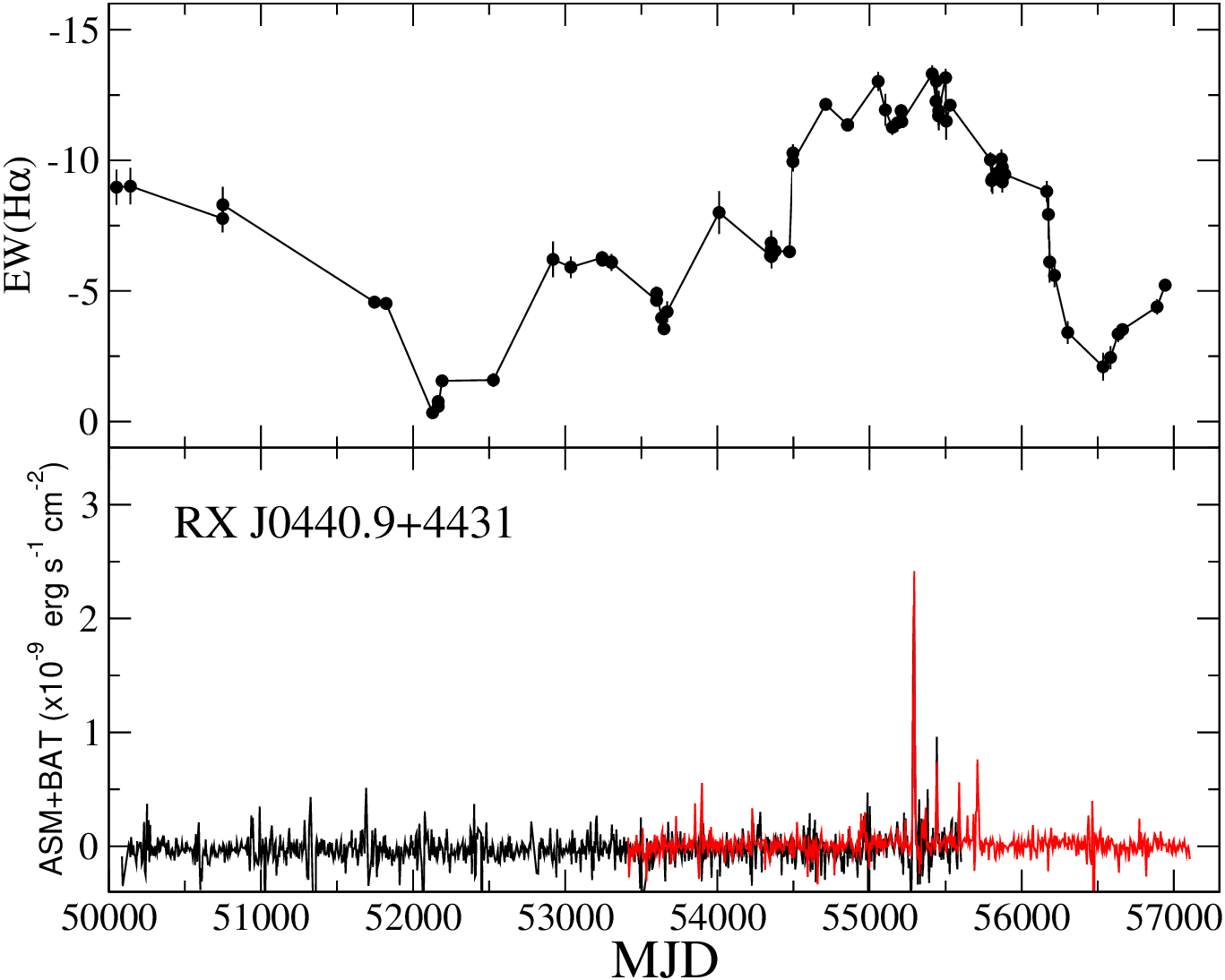} &
\includegraphics[width=7cm]{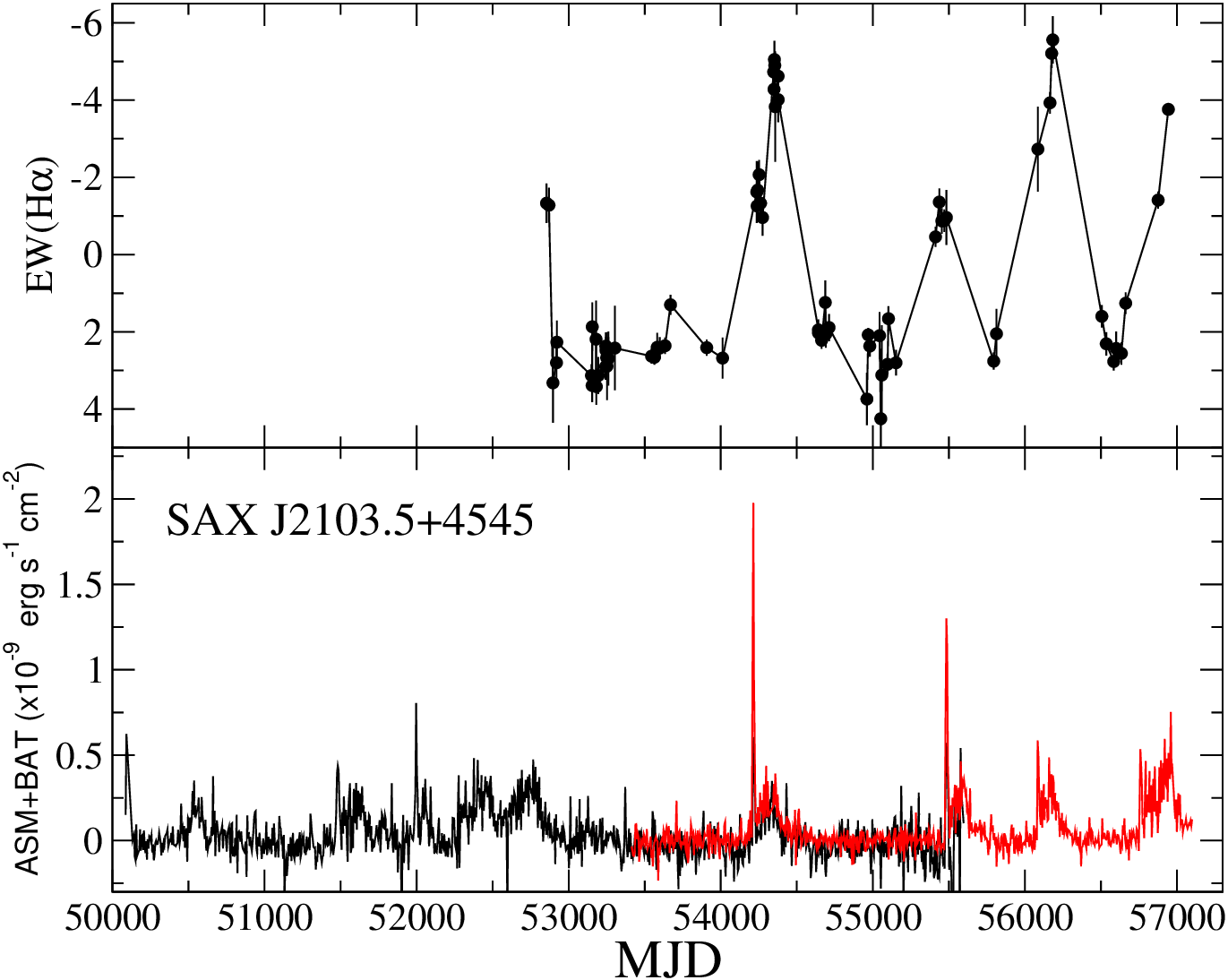} \\
\includegraphics[width=7cm]{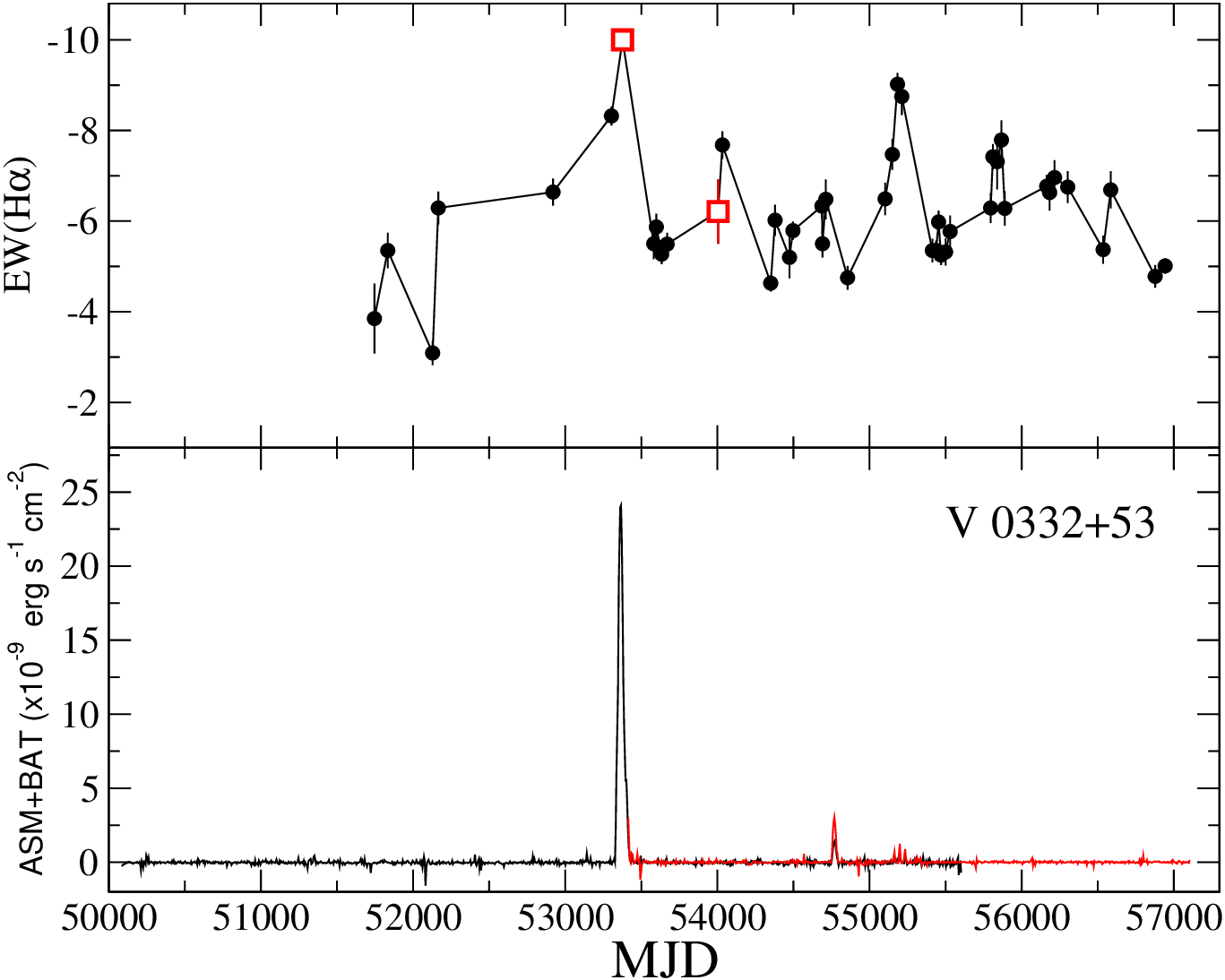} &
\includegraphics[width=7cm]{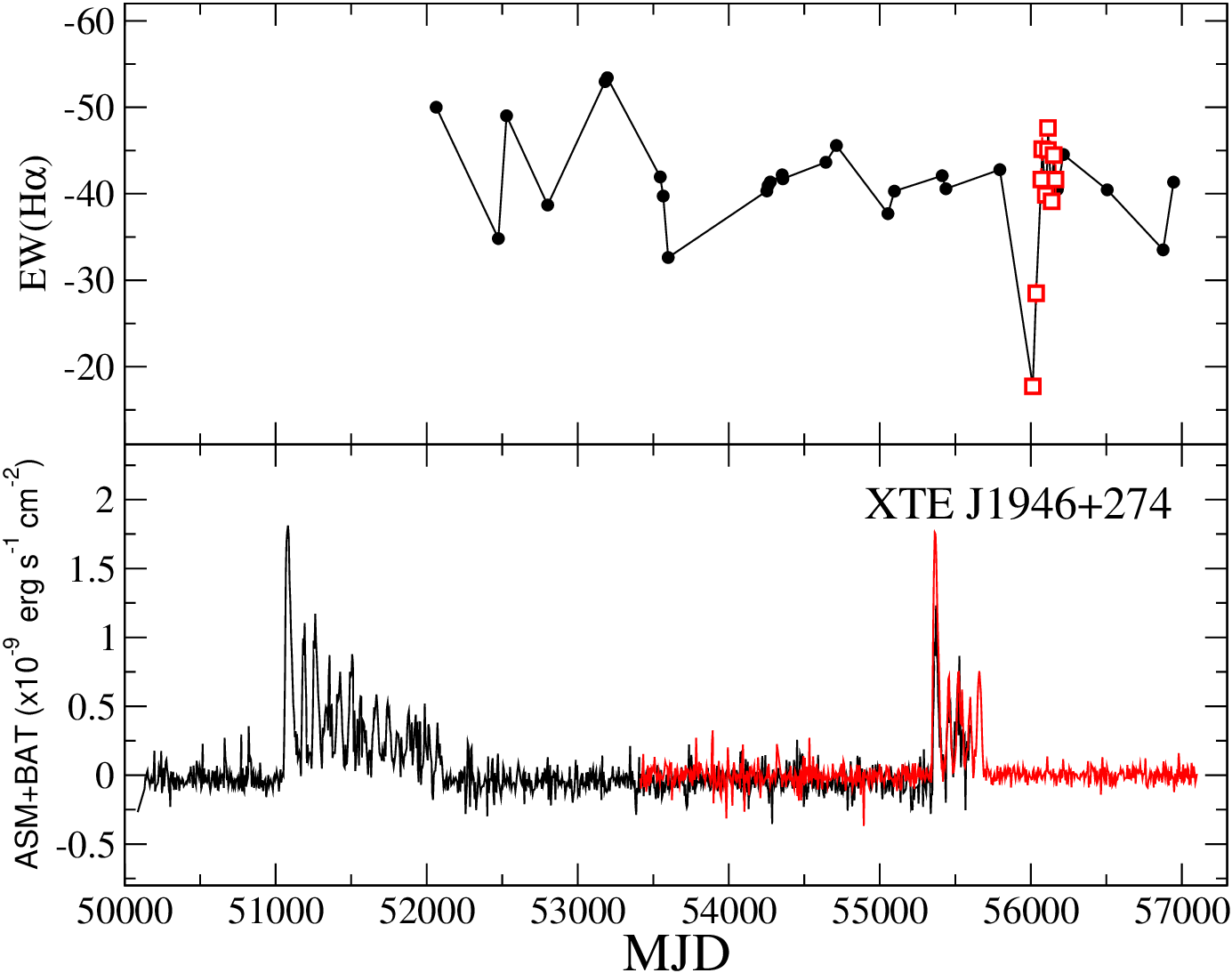} \\
\includegraphics[width=7cm]{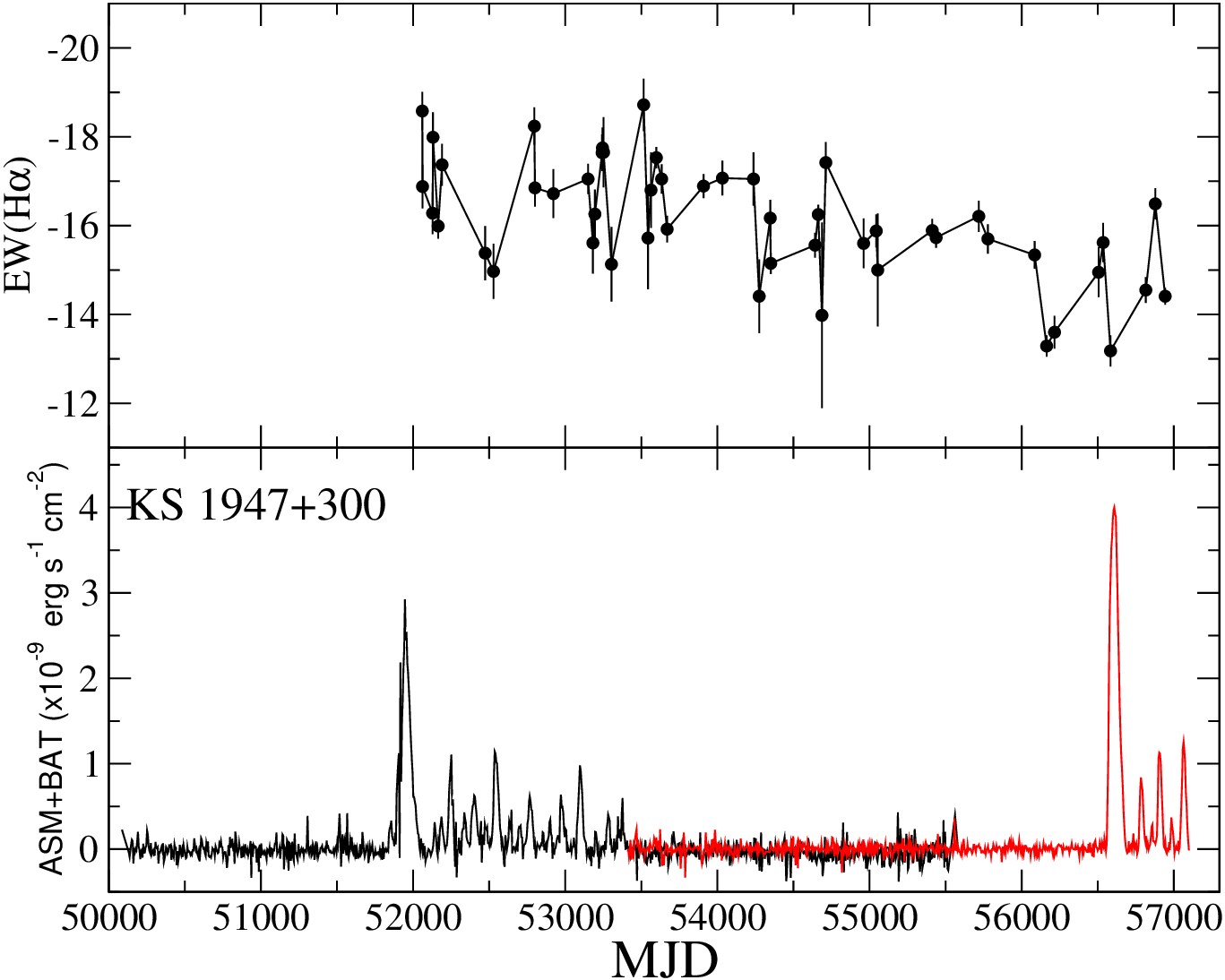} & 
\includegraphics[width=7cm]{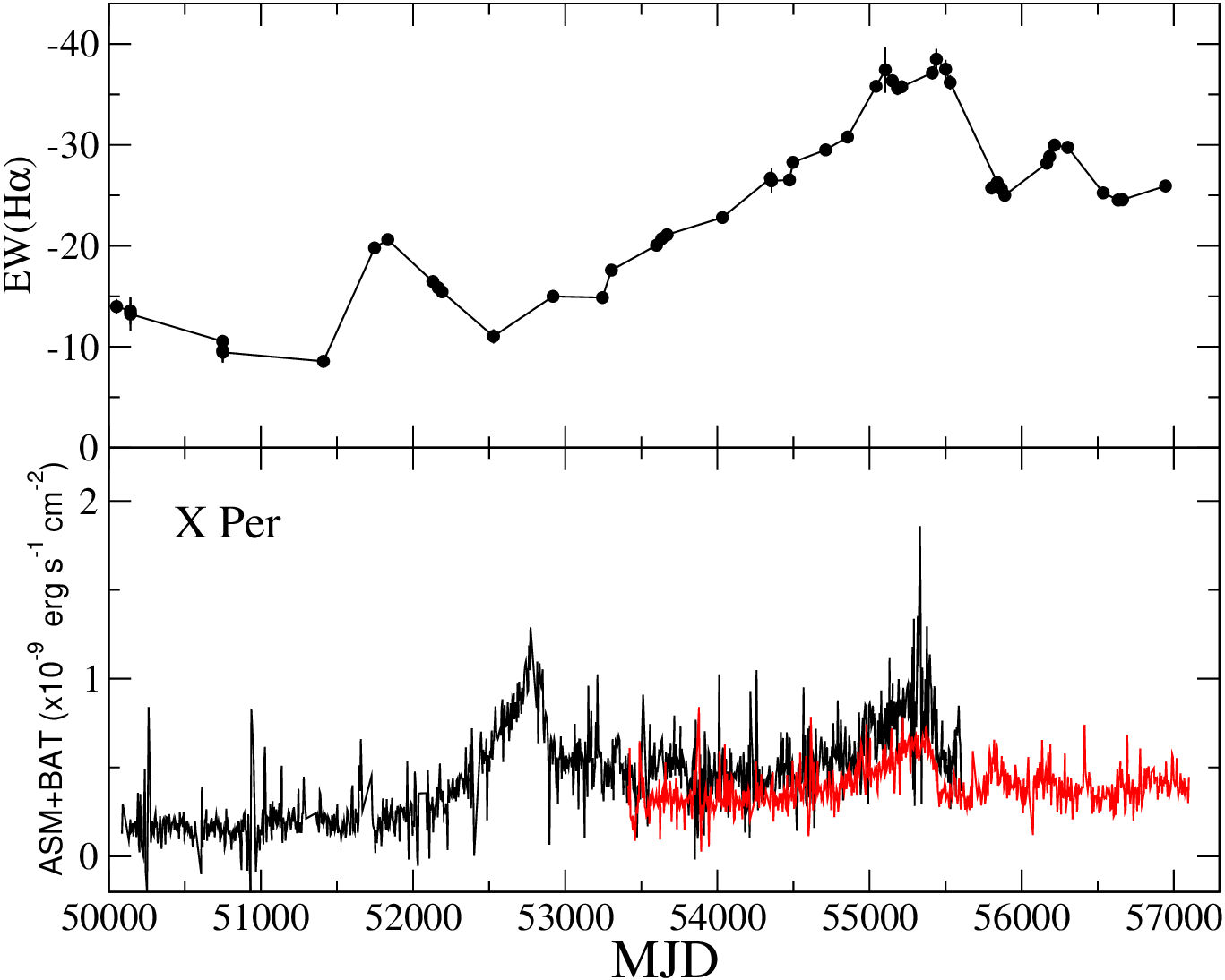} \\ 
\end{tabular}
\end{center}
\caption[]{\ha\ equivalent width and X-ray variability. The light curves
of the {\em RXTE}/ASM (2--10 keV) in black and the {\em Swift}/BAT (15-50 keV)
in red  were rebinned to 5-day bins. }
\label{ew-X}
\end{figure*}

Figure~\ref{ew-X} shows the relationship between \ew\ and the X-ray intensity
for all the sources in our list that display significant X-ray variability
during the time span covered by our optical observations. The lower panels 
shows the long-term X-ray light curve from the all-sky monitors on board
{\it RXTE} (ASM) and {\it Swift} (BAT)\footnote{Count rates were converted
into flux  by taking into account the count rate to flux conversion factors
of the two instruments (1 mCrab corresponds to 0.075
ASM count s$^{-1}$ and to 0.00022 BAT count cm$^{-2}$ s$^{-1}$) and
assuming that 1 Crab equals  $\sim2.4 \times 10^{-8}$ erg s$^{-1}$
cm$^{-2}$ in the 2--10 keV and $\sim1.3 \times 10^{-8}$ erg s$^{-1}$
cm$^{-2}$ in the 15--50 keV band. The conversion factors were taken from 
http://www.dsf.unica.it/\~riggio/calcs.html, based on \citet{kirsch05}}.

Because the X-rays are powered by matter transferred from the disk into the
neutron star, we would expect to observe a smaller disk, i.e. a
decrease of the \ew\ after X-ray outbursts. 
This expected behaviour is indeed seen in half of the systems (Fig.~\ref{ew-X}).
In RX\,J0440.9+4431, 1A\,0535+262,  SAX\,J2103.5+4545, and  4U\,0115+63, we
observe that X-ray outbursts occur when the disk radius is large, i.e.
when \ew\ is close to a local maximum. We also see a significant decline
after the  outburst. We note that in some cases (4U\,0115+63 and
SAX\,J2103.5+4545), the X-ray outburst is such a dramatic event that
disrupts the disk completely and leads to its loss, as implied by the fact
that \ew\ becomes positive.

However, some other systems do not show such a clear correlation. The X-ray
activity in  V\,0332+53, XTE\,J1946+274, and KS\,1947+300 does not seem to
have any significant effect on the \ew, if one considers the
data from this work (filled circles). In particular, the X-ray outburst observed 
in V\,0332+53 in December 2004 is one of the brightest events detected in a
BeXB. The peak luminosity reached $3.5\times10^{38}$ erg s$^{-1}$
\citep{tsygankov10} twice the value of the Eddington limit for a typical
1.4$\msun$ neutron star \citep{reig13}. Surprisingly, \ew\ did not vary
substantially. Similar behaviour is seen in XTE\,J1946+274 and 
KS\,1947+300, although in these two sources the X-ray event
was not so bright ($L_{\rm X}< 10^{38}$ erg s$^{-1}$).

One possible reason that we do not observe a significant decrease of \ew\
after the X-ray outburst could be attributed to observational gaps, i.e.  we cannot be sure that our data before and after the X-ray outburst
represent the real maximum and minimum value of the \ew.  The peak of the
outburst in V\,0332+53 occurred on MJD 53365, but our largest value of \ew\
before the outburst is --8.3 \AA\ at MJD 53304, while after the outburst
our first value is at MJD 53582 and the minimum is found at MJD 53670 with
--5.4 \AA. However, we do not know whether \ew\ was larger before MJD 53304
and smaller after MJD 53670. To investigate this issue, we searched the
literature for other measurements of \ew\ in the interval MJD 52920--54033
and found two values at MJD 53378 \citep{Masetti05} and MJD 54005
\citep{caballero15} with \ha\ equivalent widths of --10 \AA\ and --6.2
\AA, respectively. These two points have been depicted in Fig.~\ref{ew-X}
with empty red squares. Although the data point from \citet{caballero15}
narrows the gap of a possible minimum after the outburst, it does not help
solve the issue of whether our data after the outburst represent the actual
minimum value achieved by the source. The measurement by \citet{Masetti05}
shows that the disk was larger, hence the decrease in \ew\ after the X-ray
outburst was more significant  than what our data alone suggested. The low
inclination of the system may also explain the smaller amplitude change in
\ew\ after the outburst with respect to other systems, as pointed out in
Sect.~\ref{varporb}.

Observational gaps may also affect our results on XTE\,J1946+274. There are no
optical observations prior to the 1998 outburst  because it was this event
that led to the discovery of the source as a BeXB. If we look at our data
(filled circles in Fig.~\ref{ew-X}), we do not see any significant change
during or after the second X-ray active phase in 2010. The last X-ray peak
occurred in April 2011 (MJD $\sim$55660). Between our observations, taken in
August 2011 (MJD 55795) and September 2012 (MJD 56176), there is a gap of
more than a year. We found ten observations (empty red squares) from
\citet{ozbey15} during the gap period, in which the equivalent width of two of them 
was significantly smaller. Therefore, the disk in XTE\,J1946+274 did indeed
weaken after the X-ray activity, although the minimum \ew\ measured after
the outburst is still large ($\sim -20$ \AA).

Although KS\,1947+300 does show a small decrease of about 10-15\%
immediately after the peak of the outburst both in the 2000 and 2013 events
and displays a long-term gradual decrease of \ew\ between the outbursts, which is in
excellent agreement with the fading of the photometric magnitudes
\citep{reig15}, the pre-outburst level is quickly recovered. We did not
find any more data after the X-ray activity phase in the literature. The
gaps are, however, shorter, hence it is unlikely that we have missed an
extended decrease of \ew. The lack of any major observational change in the
disk of KS\,1947+300 after a giant X-ray outburst might be attributed to
its circular orbit, as explained in Sect.~\ref{varporb}. We note that, owing to
its circular orbit, the radius of the disk must reach the orbital
separation ($a$) to produce an X-ray outburst. For $P_{\rm orb}=40$ d,
$M_{*}=17.5 \msun$, and $R_*=7.7 \rsun$ this implies $a\approx 9 \times
10^{12} {\rm cm} \sim 17\,R_*$. 

The same arguments can be used for X\,Per. This system also has  a
low-eccentric orbit ($e=0.1$) and shows no correlation between optical and
X-ray emission. The difference with KS\,1947+300 is the much longer orbital
period of 250 days (hence the gap between the disk's outer radius and the
periastron distance is even wider than in KS\,1947+300)  and the fact
that X\,Per is a persistent X-ray source. X\,Per does not show the typical
Type I or II X-ray outbursts, but shows extended periods of low and high X-ray
activity. The lack of a correlation between optical and X-ray flux in
X\,per has been reported before \citep{lutovinov12,hui14}.

Another explanation for the lack of a large amplitude change in \ew\ after
major X-ray outbursts is the fact that the amount of mass accreted to
produce the increased X-ray flux represents a small part of the total mass
of the disk. KS\,1947+300, XTE\,J1946+274, and X\,Per share in common a
relatively large separation between the two components of the binary, which is either
due to the long orbital period or the low eccentricity of the orbit. As
explained before, this implies the possibility of developing very large
disks.  Following \citet{reig15}\footnote{Note that in \citet{reig15} there
is an erratum, $\dot{M}$ should read $5.4\times 10^{16}$ g s$^{-1}$ instead
of $2.7\times 10^{17}$ g s$^{-1}$. This does not affect the conclusion stated
therein because the subsequent calculations are correct.}, and assuming a
disk radius $R_{\rm disk}=8 R_*$, we estimate that  less
than 20\% of the material in the disk needs be accreted to produce the kind
of outburst ($L_X\sim 1\times 10^{37}$ for about 100 days) observed in 
KS\,1947+300 or XTE\,J1946+274. 

\section{Discussion}
\label{discussion}

Be stars may exist as isolated systems (classical Be stars) or be part of
X-ray binaries (BeXBs), where the companion is a neutron star. In both
cases, the circumstellar disk is responsible for their long-term spectral
variability. Given the large mass ratio between the two components in a
BeXB, the neutron star exerts a negligible effect on the underlying massive
companion. Hence it is normally assumed that the physical properties (mass,
radius, and luminosity) of the B star in an isolated Be star and in a BeXB are
the same for the same spectral type. However, the circumstellar disk may
extend over several stellar radii and occasionally reach periastron
distance. It is sensible to think that the presence of a second body (the
neutron star) must have some effect on the observational properties of the
disk. As the equatorial disk is more exposed to the action of the neutron
star, we would expect to see different properties in the disks of isolated
Be stars compared to those of BeXBs. The most important difference is the
truncation of the disk 
\citep{reig97a,negueruela01a,okaneg01,okazaki01,okazaki02,reig11}. 

If there is no physical difference between the B star in an isolated Be
star and the B star in a BeXB, and the optical spectral variability is
driven by changes in the disk, then any difference in the observational
properties between classical Be stars and BeXBs must come from the effect of the
neutron star on the disk. In this work, we searched for these types of differences in
the variability patterns of various physical parameters and found important
differences in disk density and size, and the characteristic variability
timescales associated with the evolution of the disk (V/R changes, disk
dissipation and reformation).  Further evidence for the interaction between
the disk and the neutron star comes from the correlation of disk parameters
and the orbital period because one would expect that the closer the two
components, the larger the strength of the interaction.

In what follows, we summarise the observational evidence for the truncation
of the disk, gathered in two groups: correlations with the orbital period
and comparison between BeXB systems and classical Be stars. Some of the
results presented here are new,  others confirm previous studies. 
However, it is worth mentioning that our work constitutes the first attempt
to explain the long-term optical variability of BeXB as a group using the
largest data set so far analysed.

\subsection{Correlation with the orbital period}

\begin{itemize}

\item The $P_{\rm orb}$--variability correlation. In sect~\ref{varcor}, we
showed that systems with small orbital periods are more variable both in
the continuum and line optical emission (Figs.~\ref{rms-ew-v}
and~\ref{rms-orb}). A similar result was found by analysing the variability
in the X-ray band \citep{reig07a}. These results strongly suggest that the
circumstellar disk in BeXBs is truncated. Because the disk of systems with
short orbital periods are more exposed to the action of the neutron star,
they cannot reach a stable configuration over long timescales.

\item Disk recovery after dissipation. In Sect.~\ref{diskgrowth}, we showed
a correlation between disk growth rate (change of \ew\ with time) and
orbital period after a low state. Systems with shorter orbital periods
display larger growth rates. Owing to truncation, the disk density increases
faster in closer binaries, leading to a faster increase in \ew.  Not only
the disk formation, but also the entire formation and dissipation cycle
appears to be faster in systems with short orbital periods, while longer
timescales are associated with longer orbital periods. This can be deduced
from Fig.~\ref{diskew}, where the systems are ordered in increasing orbital
period.  Disk-loss phases can be easily identified as intervals in which
\ew\ $>$ 0.


\item Orbital separation and disk size.  In Sect.~\ref{ewper}, we showed a
correlation between the periastron distance and the maximum measured value
of \ew. This relationship is similar to the one between the orbital period
and \ew\ \citep{reig97a,reig07a,antoniou09} or between the  orbital
separation and the size of the circumstellar disk \citep{coe15}.    In all
these diagrams, the larger the orbital size (i.e. orbital period,
periastron or distance between the components), the larger the \ew. Since
the equivalent width is directly related to the size of the disk (see
Sect.~\ref{diskradius}), these correlations imply that large disks can only
develop in systems in which the two components are well apart. In systems
with small orbital separation, the  tidal torque exerted by the
neutron star prevents the disk from expanding. 

\end{itemize}

\subsection{BeXB versus classical Be stars}

\begin{itemize}

\item On average, BeXBs have denser disks. In Sect.~\ref{diskvelaw}  (see
also \citet{reig00,zamanov01}), we showed that the disks in BeXB are at
least 1.5 times denser than disks in isolated Be stars. These results find
theoretical confirmation in the work by \citet{okazaki02}.  Using a 3D,
smoothed particle hydrodynamics code \citet{okazaki02} studied the viscous
effect on the interaction between the Be star's disk and the neutron star
and find that, owing to truncation, the material decreted from the Be star
accumulates and the disk becomes denser more rapidly than in isolated Be
stars.

\item On average, BeXBs have  smaller disks. In Sects.~\ref{hist} and
\ref{diskradius} (see also \citealt{reig97a}), we showed that, on average,
classical Be stars have larger \ha\ equivalent widths, hence larger disks. 
A Kolmogorov-Smirnov test of the two populations gives a probability of 
3.4\% that they are drawn from the same population. A similar result was found between the isolated Be stars and BeXBs in the Small Magellanic
Cloud, with a probability of 0.9\% that they are drawn from the same
population \citep{coe15}.

\item BeXBs have shorter V/R quasiperiods. The period of long-term V/R
variations is typically in the range 5-10 yr for isolated Be stars, with a
statistical average of $\sim 7$ years \citep{okazaki91,okazaki97}. In
contrast, the V/R  quasiperiods of BeXB are significantly shorter ($<5$
yr). \citet{oktariani09} find that the one-armed oscillations are
well confined in systems with disks larger than a few tens of stellar
radii. In such systems, the oscillation period scarcely depends  on the binary
parameters. On the other hand, in systems with smaller disks, where mode
confinement is incomplete, the oscillation period increases with increasing
orbital separation and/or decreasing binary mass ratio. The trend of
shorter periods in smaller systems is mainly due to the fact that the tidal
field is stronger in smaller systems. The difference in V/R quasiperiods
between classical Be stars and BeXBs can be explained by the smaller size
of the Be star's disks in BeXB systems (see Sect.~\ref{diskradius}). 

\end{itemize}

\section{Conclusion}

We have investigated the long-term optical spectroscopic variability of the
counterparts to high-mass X-ray binaries visible from the northern
hemisphere. Our work represents the most complete optical spectroscopic
study on the global properties of HMXBs, not only because of the number of
systems included, but also because of the long monitoring baseline, cadence,
and large amount of data analysed (1 164 spectra of 20 sources). 

We have focused on the variability  of the \ha\ line since this feature is
closely related to the conditions in the equatorial disk.  We have shown
that, as in classical Be stars, the disk in BeXB follows a Keplerian
velocity distribution. The radius of the disk correlates with the \ha\ line
equivalent width. Double peak profiles are found when the equivalent width
is small, indicating a small disk.  We have provided evidence for the
truncation of the disk by the neutron star. This evidence is based on the
correlation between variability timescales and the orbital period and on
the different properties of disks in isolated Be stars and BeXBs.

We conclude that the interaction between the compact object and the Be-type
star works in two directions: the massive companion provides the source of
matter for accretion; the amount of matter captured and the way it is
captured (transfer mechanism) change the physical conditions of the neutron
star (e.g. by spinning it up or down and shrinking or expanding the
magnetosphere). But also, the continuous revolution of the neutron star
around the optical counterpart produces observable effects, the most
important of which is the truncation of the Be star's equatorial disk.

Progress in understanding the interaction between the neutron star and the
circumstellar disk requires long-term monitoring campaigns to probe the
disk variability timescales (months to years). Small and medium size
telescopes are ideal for this type of study.

\begin{acknowledgements}

Skinakas Observatory is a collaborative project of the University of Crete
and the Foundation for Research and Technology-Hellas.  This paper  uses
observations made at the South African Astronomical Observatory (SAAO) and
data products produced by the OIR Telescope Data Center, supported by the
Smithsonian Astrophysical Observatory. We thank observers P. Berlind and M.
Calkins for performing the FLWO observations.  This work was supported in
part by the European Union Marie Curie grant MTKD-CT-2006-039965 and EU FP7
Capacities GA No206469. AZ acknowledges funding from the European
Research Council under the European Union's Seventh Framework Programme
(FP/2007-2013)/ERC Grant Agreement no. 617001. During the period of this
work  AZ has also been supported by the Chandra grant GO3-14051X, NASA ADAP
grant NNX12AN05G, NASA LTSA grant NAG5-13056, and the EU IRG grant 224878.
This work used NASA's Astrophysics Data System Bibliographic Services and
of the SIMBAD database, operated at the CDS, Strasbourg, France. The  WHT
and its service programme (service proposal references SW2012b14 and
SW2013a19) are operated on the island of La Palma by the Isaac Newton Group
in the Spanish Observatorio del Roque de los Muchachos of the Instituto de
Astrof\'{\i}sica de Canarias. Swift/BAT transient monitor results provided
by the Swift/BAT team. Quick-look results provided by the ASM/RXTE team.


\end{acknowledgements}

\bibliographystyle{aa}
\bibliography{./artBex_bib}

\clearpage

\begin{appendix}

\section{Data}

The following tables provide the equivalent width of the \ha\ line, the
type of the \ha\ profile, and the telescope used for the sources analysed in
this work.   Our optical spectroscopic monitoring programme of BeXB systems
started in 1999 using the 1.3 m telescope of the Skinakas Observatory.
During the period  2007-2014, the targets were also regularly observed from
the 1.5 m of the Fred Lawrence Whipple Observatory.   The star symbol in
some of the observations taken from SKO and FLWO indicates that the grating
with 2400 l mm$^{-1}$ and 600 l mm$^{-1}$ was used instead of the default
set-up of  1302 l mm$^{-1}$ and 1200 l mm$^{-1}$ gratings, respectively (see
Sect.~\ref{observations}). The identification of the line profile is as
follows (see Sect.~\ref{profvar}): absorption (AB), single-peak (SP),
fill-in (FI), double-peak (DP), multi-peak (MP), shell (SH), and P Cygni
(PC).

Although the observations from Skinakas and Fred Lawrence Whipple
observatories represent the main bulk of data used in this work, we 
complemented these data with older observations made in the framework of
the Southampton/Valencia collaboration. Those observations were made with
a variety of telescopes as indicated in the tables.

The identification of the telescopes is as follows: 
1.3 m Skinakas Observatory (SKO), 
1.5 m Fred Lawrence Wripple Observatory (FLWO), 
4.2 m William Herschel Telescope  (WHT), 
2.5 m Isaac Newton Telescope (INT), 
1.0 m Jacobus Kapteyn Telescope (JKT), 
2.5 m Nordic Optical Telescope (NOT),
1.9 m South Africa Astronomical Observatory (SAAO),
1.5 m at Palomar Mountain (PAL), and
2.2 m telescope at Calar Alto (CAL),
2.1 m telescope of Kitt Peak National Observatory (KPNO).

\begin{table}
\caption{Results for 2S\,0114+65}
\label{2s0114}
\begin{center}
\begin{tabular}{l@{~}c@{~}c@{~~~~~~}l@{~~~~}l}
\hline  \hline
Date            &JD            &\ew\          &Profile          &Telescope \\
                &(2,4000,000+) &(\AA)         &                 &        \\     
\hline\hline
26/12/90        &48252.500     &-2.14$\pm$0.35    &MP      &INT     \\
27/12/90        &48253.500     &-2.06$\pm$0.38    &MP      &INT     \\
27/01/91        &48284.500     &-1.92$\pm$0.38    &PC      &INT     \\
28/08/91        &48497.500     &-1.40$\pm$0.31    &MP      &INT     \\
13/12/91        &48604.500     &-1.88$\pm$0.46    &SP      &INT     \\
18/02/92        &48671.500     &-1.98$\pm$0.39    &SP      &PAL     \\
\hline\hline
\end{tabular}
\end{center}
\end{table}

\clearpage

\begin{table}
\caption{Results for 4U\,0115+63}
\label{4u0115}
\begin{center}
\begin{tabular}{l@{~}c@{~}c@{~~~~~~}l@{~~~~}l}
\hline  \hline
Date            &JD            &\ew\          &Profile  &Telescope \\
                &(2,4000,000+) &(\AA)         &         &        \\     
\hline\hline
14/02/1990      &47937.5       &-6.70$\pm$0.84    &DP   &WHT    \\
27/12/1990      &48253.5       &-3.53$\pm$0.54    &DP   &INT     \\
27/01/1991      &48284.5       &-8.97$\pm$0.79    &DP   &INT     \\
28/08/1991      &48497.5       &-4.92$\pm$1.13    &SH   &INT     \\
14/12/1991      &48605.5       &-5.33$\pm$1.05    &DP   &INT     \\
05/08/1992      &48840.5       &-3.72$\pm$0.41    &DP   &INT     \\
\hline\hline
\end{tabular}
\end{center}
\end{table}

\begin{table}
\caption{Results for 4U\,0115+63 (cont.)}
\label{4u0115c}
\begin{center}
\begin{tabular}{l@{~}c@{~}c@{~~~~~~}l@{~~~~}l}
\hline  \hline
Date            &JD            &\ew\          &Profile  &Telescope \\
                &(2,4000,000+) &(\AA)         &         &        \\     
\hline\hline
17/08/2005      &53600.5       &-16.42$\pm$1.49   &DP   &SKO     \\
18/08/2005      &53601.5       &-17.09$\pm$1.81   &DP   &SKO     \\
20/09/2005      &53634.5       &-13.79$\pm$1.63   &DP   &SKO     \\
26/10/2005      &53670.5       &-5.51$\pm$0.79    &DP   &SKO     \\
20/06/2006      &53907.5       &+0.61$\pm$0.78    &SH   &SKO     \\
21/06/2006      &53908.5       &+0.80$\pm$0.48    &FI   &SKO     \\
\hline\hline
\end{tabular}
\end{center}
\end{table}

\clearpage

\begin{table}
\caption{Results for IGR\,J01363+6610}
\label{igr01360}
\begin{center}
\begin{tabular}{l@{~}c@{~}c@{~~~~~~}l@{~~~~}l}
\hline  \hline
Date            &JD            &\ew\          &Profile  &Telescope \\
                &(2,4000,000+) &(\AA)         &         &       \\     
\hline\hline
06/07/2004      &53193.5       &-52.93$\pm$0.94    &SP  &SKO        \\
26/08/2004      &53244.5       &-50.73$\pm$0.59    &SP  &SKO      \\
12/09/2004      &53261.5       &-51.46$\pm$0.55    &SP  &SKO      \\
13/09/2004      &53262.5       &-46.10$\pm$2.63    &SP  &SKO      \\
03/10/2004      &53282.5       &-50.51$\pm$0.59    &SP  &SKO      \\
24/10/2004      &53303.5       &-50.57$\pm$0.47    &SP  &SKO      \\
\hline\hline
\end{tabular}
\end{center}
\end{table}

\clearpage

\begin{table}
\caption{Results for RX\,J0146.9+6121}
\label{lsi61235}
\begin{center}
\begin{tabular}{l@{~}c@{~}c@{~~~~~~}l@{~~~~}l}
\hline  \hline
Date            &JD            &\ew\          &Profile  &Telescope \\
                &(2,4000,000+) &(\AA)         &         &      \\     
\hline\hline
28/08/1991      &48497.5       &-8.40 $\pm$0.79    &DP      &INT     \\
13/12/1991      &48604.5       &-6.94 $\pm$0.27    &DP      &INT     \\
18/05/1992      &48761.5       &-6.18 $\pm$1.38    &SH      &PAL     \\
19/05/1992      &48762.5       &-6.33 $\pm$1.41    &SH      &PAL     \\
16/08/1992      &48851.5       &-7.82 $\pm$1.08    &DP      &PAL     \\
17/08/1992      &48852.5       &-8.49 $\pm$0.74    &DP      &PAL     \\
\hline\hline
\end{tabular}
\end{center}
\end{table}

\begin{table}
\caption{Results for RX\,J0146.9+6121 (cont.)}
\label{lsi61235c}
\begin{center}
\begin{tabular}{l@{~}c@{~}c@{~~~~~~}l@{~~~~}l}
\hline  \hline
Date            &JD            &\ew\          &Profile  &Telescope \\
                &(2,4000,000+) &(\AA)         &         &      \\     
\hline\hline
29/07/2005      &53581.5       &-3.97 $\pm$0.32    &DP      &SKO     \\
17/08/2005      &53600.5       &-3.74 $\pm$0.14    &DP      &SKO     \\
20/09/2005      &53634.5       &-3.82 $\pm$0.19    &DP      &SKO     \\
26/10/2005      &53670.5       &-4.26 $\pm$0.22    &DP      &SKO     \\
02/10/2006      &54011.5       &-4.52 $\pm$0.40    &DP      &SKO     \\
04/09/2007      &54348.5       &-5.54 $\pm$0.16    &DP      &SKO     \\
\hline\hline
\end{tabular}
\end{center}
\end{table}
\clearpage

\begin{table}
\caption{Results for IGR\,J01583+6713}
\label{igr01583}
\begin{center}
\begin{tabular}{l@{~}c@{~}c@{~~~~~~}l@{~~~~}l}
\hline  \hline
Date            &JD            &\ew\          &Profile  &Telescope \\
                &(2,4000,000+) &(\AA)         &         &      \\     
\hline\hline
24/10/2006      &54033.5       &-68.44$\pm$0.82    &SP  &SKO       \\
29/05/2007      &54250.5       &-66.92$\pm$0.88    &SP  &SKO      \\
04/09/2007      &54348.5       &-59.04$\pm$2.52    &SP  &SKO      \\
07/09/2007      &54351.5       &-63.94$\pm$1.44    &SP  &SKO      \\
12/09/2007      &54356.5       &-64.67$\pm$1.14    &DP  &FLWO     \\
02/10/2007      &54376.5       &-63.97$\pm$1.00    &SP  &SKO      \\
\hline\hline
\end{tabular}
\end{center}
\end{table}

\clearpage

\begin{table}
\caption{Results for RX\,J0240.4+6112}
\label{lsi61303}
\begin{center}
\begin{tabular}{l@{~}c@{~}c@{~~~~~~}l@{~~~~}l}
\hline  \hline
Date            &JD            &\ew\          &Profile \\
                &(2,4000,000+) &(\AA)         &        \\     
\hline\hline
27/12/1990      &48253.5       &-10.95$\pm$0.95    &DP  &INT     \\
27/01/1991      &48284.5       &-13.71$\pm$0.82    &DP  &INT     \\
28/08/1991      &48497.5       &-10.85$\pm$0.89    &DP  &INT     \\
13/12/1991      &48604.5       &-11.08$\pm$1.45    &DP  &INT     \\
16/08/1992      &48851.5       &-15.78$\pm$2.06    &DP  &PAL     \\
17/08/1992      &48852.5       &-15.51$\pm$1.10    &DP  &PAL     \\
\hline\hline
\end{tabular}
\end{center}
\end{table}

\begin{table}
\caption{Results for RX\,J0240.4+6112 (cont.)}
\label{lsi61303c}
\begin{center}
\begin{tabular}{l@{~}c@{~}c@{~~~~~~}l@{~~~~}l}
\hline  \hline
Date            &JD            &\ew\          &Profile \\
                &(2,4000,000+) &(\AA)         &        \\     
\hline\hline
03/11/2010      &55504.5       &-13.03$\pm$0.59    &DP  &SKO      \\
29/11/2010      &55530.5       &-11.57$\pm$0.40    &DP  &FLWO     \\
21/08/2011      &55795.5       &-12.86$\pm$0.66    &DP  &SKO      \\
06/09/2011      &55811.5       &-12.82$\pm$0.77    &DP  &SKO*     \\
03/10/2011      &55838.5       &-11.57$\pm$0.82    &DP  &FLWO     \\
01/11/2011      &55867.5       &-10.20$\pm$0.28    &DP  &FLWO     \\
23/11/2011      &55889.5       &-11.01$\pm$0.27    &DP  &FLWO     \\
06/09/2012      &56177.5       &-15.90$\pm$1.46    &DP  &SKO      \\
15/10/2012      &56216.5       &-12.69$\pm$0.28    &DP  &FLWO     \\
11/01/2013      &56304.5       & -9.42$\pm$0.59    &DP  &FLWO     \\
31/07/2013      &56505.5       &-15.03$\pm$1.21    &DP  &SKO      \\
06/08/2014      &56876.5       & -8.95$\pm$0.49    &DP  &SKO      \\
14/10/2014      &56945.5       &-11.14$\pm$0.67    &DP  &SKO      \\
\hline\hline
\end{tabular}
\end{center}
\end{table}

\clearpage

\begin{table}
\caption{Results for V\,0332+53}
\label{v0332}
\begin{center}
\begin{tabular}{l@{~}c@{~}c@{~~~~~~}l@{~~~~}l}
\hline  \hline
Date            &JD            &\ew\          &Profile  &Telescope \\
                &(2,4000,000+) &(\AA)         &         &      \\     
\hline\hline
28/01/1990      &47920.5       &-5.38$\pm$0.66    &DP   &INT    \\
02/09/1990      &48137.5       &-5.91$\pm$0.43    &DP   &INT     \\
13/11/1990      &48209.5       &-5.90$\pm$0.92    &DP   &INT     \\
14/11/1990      &48210.5       &-6.09$\pm$0.30    &SP   &INT     \\
27/12/1990      &48253.5       &-7.68$\pm$1.64    &SP   &INT     \\
27/01/1991      &48284.5       &-6.59$\pm$0.40    &SP   &INT     \\
\hline\hline
\end{tabular}
\end{center}
\end{table}

\clearpage

\begin{table}
\caption{Results for X\,Per}
\label{xper}
\begin{center}
\begin{tabular}{l@{~}c@{~}c@{~~~~~~}l@{~~~~}l}
\hline  \hline
Date            &JD            &\ew\          &Profile  &Telescope \\
                &(2,4000,000+) &(\AA)         &         &   \\     
\hline\hline    
24/12/1987      &47154.5       &-7.13$\pm$2.05    &SP   &JKT     \\
31/12/1987      &47161.5       &-9.68$\pm$0.91    &SP   &JKT     \\
15/11/1990      &48211.5       &+2.59$\pm$0.33    &AB   &INT     \\
27/12/1990      &48253.5       &+1.92$\pm$0.56    &AB   &INT     \\
27/01/1991      &48284.5       &+2.17$\pm$0.67    &AB   &INT     \\
28/08/1991      &48497.5       &+2.18$\pm$1.23    &AB   &INT     \\
\hline\hline
\end{tabular}
\end{center}
\end{table}

\begin{table}
\caption{Results for X\,Per (cont.)}
\label{xperc}
\begin{center}
\begin{tabular}{l@{~}c@{~}c@{~~~~~~}l@{~~~~}l}
\hline  \hline
Date            &JD            &\ew\          &Profile  &Telescope \\
                &(2,4000,000+) &(\AA)         &         &   \\     
\hline\hline    
18/12/2009      &55184.5       &-35.61$\pm$0.67    &SP  &FLWO*    \\
16/01/2010      &55213.5       &-35.76$\pm$0.49    &SP  &FLWO*    \\
03/08/2010      &55412.5       &-37.14$\pm$0.65    &SP  &SKO      \\
29/08/2010      &55438.5       &-38.49$\pm$1.02    &SP  &SKO      \\
30/10/2010      &55500.5       &-37.50$\pm$0.92    &SP  &FLWO     \\
29/11/2010      &55530.5       &-36.18$\pm$0.73    &SP  &FLWO     \\
03/10/2011      &55838.5       &-26.28$\pm$0.52    &SP  &FLWO     \\
28/08/2011      &55802.5       &-25.73$\pm$0.64    &SP  &SKO      \\
01/11/2011      &55867.5       &-25.63$\pm$0.38    &SP  &FLWO     \\
23/11/2011      &55889.5       &-25.00$\pm$0.42    &SP  &FLWO     \\
15/10/2012      &56216.5       &-29.98$\pm$0.50    &SP  &FLWO     \\
25/08/2012      &56165.5       &-28.18$\pm$0.44    &SP  &SKO      \\
14/09/2012      &56185.5       &-28.84$\pm$0.81    &SP  &SKO      \\
11/01/2013      &56304.5       &-29.76$\pm$0.61    &SP  &FLWO     \\
31/08/2013      &56536.5       &-25.24$\pm$0.42    &SP  &SKO      \\
07/12/2013      &56634.5       &-24.53$\pm$0.27    &SP  &FLWO     \\
06/01/2014      &56664.5       &-24.56$\pm$0.33    &SP  &FLWO     \\
14/10/2014      &56945.5       &-25.93$\pm$0.12    &SP  &SKO      \\
\hline\hline
\end{tabular}
\end{center}
\end{table}

\clearpage

\begin{table}
\caption{Results for RX\,J0440.9+4431}
\label{ls4417}
\begin{center}
\begin{tabular}{l@{~}c@{~}c@{~~~~~~}l@{~~~~}l}
\hline  \hline
Date            &JD            &\ew\          &Profile  &Telescope \\
                &(2,4000,000+) &(\AA)         &         &  \\     
\hline\hline
29/11/1995      &50051.5       &-8.97$\pm$0.67    &DP   &JKT     \\
28/02/1996      &50142.5       &-9.01$\pm$0.70    &DP   &JKT     \\
26/10/1997      &50748.5       &-7.77$\pm$0.53    &DP   &JKT     \\
28/10/1997      &50750.5       &-8.30$\pm$0.68    &DP   &JKT     \\
20/07/2000      &51746.5       &-4.57$\pm$0.23    &DP   &SKO     \\
05/10/2000      &51823.5       &-4.52$\pm$0.21    &DP   &SKO     \\
\hline\hline
\end{tabular}
\end{center}
\end{table}

\begin{table}
\caption{Results for RX\,J0440.9+4431 (cont.)}
\label{ls4417c}
\begin{center}
\begin{tabular}{l@{~}c@{~}c@{~~~~~~}l@{~~~~}l}
\hline  \hline
Date            &JD            &\ew\          &Profile  &Telescope \\
                &(2,4000,000+) &(\AA)         &         &  \\     
\hline\hline
15/10/2012      &56216.5       &-5.59$\pm$0.45    &DP   &FLWO    \\
11/01/2013      &56304.5       &-3.41$\pm$0.43    &DP   &FLWO    \\
30/08/2013      &56535.5       &-2.10$\pm$0.53    &DP   &SKO     \\
18/10/2013      &56584.5       &-2.45$\pm$0.44    &DP   &SKO     \\
07/12/2013      &56634.5       &-3.35$\pm$0.31    &DP   &FLWO    \\
06/01/2014      &56664.5       &-3.52$\pm$0.19    &DP   &FLWO    \\
19/08/2014      &56889.5       &-4.39$\pm$0.29    &DP   &SKO     \\
12/10/2014      &56943.5       &-5.22$\pm$0.23    &DP   &SKO     \\
\hline\hline
\end{tabular}
\end{center}
\end{table}

\clearpage

\begin{table}
\caption{Results for 1A\,0535+262}
\label{a0535}
\begin{center}
\begin{tabular}{l@{~}c@{~}c@{~~~~~~}l@{~~~~}l}
\hline  \hline
Date            &JD            &\ew\          &Profile  &Telescope \\
                &(2,4000,000+) &(\AA)         &         &       \\     
\hline\hline
24/12/1987      &47154.5       &-8.23$\pm$0.55    &DP   &INT            \\
31/12/1987      &47161.5       &-8.04$\pm$0.36    &DP   &INT    \\
14/11/1990      &48210.5       &-9.23$\pm$0.26    &DP   &INT    \\
27/12/1990      &48253.5       &-9.47$\pm$1.16    &DP   &INT    \\
27/01/1991      &48284.5       &-9.49$\pm$0.27    &DP   &INT    \\
16/04/1991      &48363.5       &-7.06$\pm$0.26    &DP   &INT    \\
\hline\hline
\end{tabular}
\end{center}
\end{table}

\begin{table}
\caption{Results for 1A\,0535+262 (cont.)}
\label{a0535c}
\begin{center}
\begin{tabular}{l@{~}c@{~}c@{~~~~~~}l@{~~~~}l}
\hline  \hline
Date            &JD            &\ew\          &Profile  &Telescope \\
                &(2,4000,000+) &(\AA)         &         & \\     
\hline\hline
08/01/2008      &54474.5       &-11.87$\pm$0.23    &DP  &FLWO             \\
31/01/2008      &54497.5       &-12.83$\pm$0.15    &SP  &FLWO    \\
02/09/2008      &54712.5       &-14.55$\pm$0.29    &SP  &SKO     \\
25/01/2009      &54857.5       &-16.04$\pm$0.27    &DP  &FLWO    \\
11/08/2009      &55055.5       &-18.84$\pm$0.57    &SP  &SKO     \\
14/08/2009      &55058.5       &-18.89$\pm$0.62    &SP  &SKO     \\
\hline\hline
\end{tabular}
\end{center}
\end{table}

\clearpage

\begin{table}
\caption{Results for IGR\,J06074+2205}
\label{igr06074}
\begin{center}
\begin{tabular}{l@{~}c@{~}c@{~~~~~~}l@{~~~~}l}
\hline  \hline
Date            &JD            &\ew\          &Profile  &Telescope \\
                &(2,4000,000+) &(\AA)         &         &       \\     
\hline\hline
22/10/2006      &54031.5       &-11.36$\pm$0.25    &SP  &SKO       \\
24/10/2006      &54033.5       &-11.41$\pm$0.11    &DP  &SKO      \\
06/09/2007      &54350.5       &-11.95$\pm$0.27    &DP  &SKO      \\
12/09/2007      &54356.5       &-12.46$\pm$0.74    &DP  &FLWO     \\
04/10/2007      &54378.5       &-10.51$\pm$0.18    &DP  &SKO      \\
08/01/2008      &54474.5       &-6.81$\pm$0.19    &DP   &FLWO*   \\
\hline\hline
\end{tabular}
\end{center}
\end{table}

\clearpage

\begin{table}
\caption{Results for AX\,J1845.0-0433}
\label{ax1845}
\begin{center}
\begin{tabular}{l@{~}c@{~}c@{~~~~~~}l@{~~~~}l}
\hline  \hline
Date            &JD            &\ew\          &Profile  &Telescope \\
                &(2,4000,000+) &(\AA)         &         &   \\     
\hline\hline
07/06/1995      &49876.5     &-2.56$\pm$0.55    &PC,MP  &INT   \\
15/07/1995      &49914.5     &-3.46$\pm$0.56    &PC,MP  &INT   \\
04/08/1995      &49934.5     &-3.95$\pm$0.65    &PC,MP  &JKT   \\
04/10/1996      &50361.5     &-4.95$\pm$1.04    &PC,MP  &SAAO  \\
05/10/1996      &50362.5     &-4.82$\pm$0.84    &PC,MP  &SAAO  \\
05/08/1998      &51031.5     &-3.19$\pm$0.74    &PC,MP  &INT   \\
\hline\hline
\end{tabular}
\end{center}
\end{table}

\clearpage

\begin{table}
\caption{Results for 4U\,1907+09}
\label{4u1907}
\begin{center}
\begin{tabular}{l@{~}c@{~}c@{~~~~~~}l@{~~~~}l}
\hline  \hline
Date            &JD            &\ew\          &Profile  &Telescope \\
                &(2,4000,000+) &(\AA)         &         &
     \\     
\hline\hline
27/08/1991      &48496.5     &-7.67$\pm$0.86	  &MP	      &INT \\
29/06/1993      &49168.5     &-6.44$\pm$1.05	  &SP$^\dag$  &WHT \\
30/06/1993      &49169.5     &-6.02$\pm$1.26	  &MP	      &WHT \\
01/07/1994      &49535.5     &-6.03$\pm$1.12	  &MP	      &SAAO\\
03/07/1994      &49537.5     &-5.33$\pm$1.50	  &MP	      &SAAO\\
10/07/1995      &49909.5     &-6.01$\pm$1.57	  &SP$^\dag$  &WHT \\
\end{tabular}
\end{center}
\end{table}

\clearpage

\begin{table}
\caption{Results for XTE\,J1946+274}
\label{xte1946}
\begin{center}
\begin{tabular}{l@{~}c@{~}c@{~~~~~~}l@{~~~~}l}
\hline  \hline
Date            &JD            &\ew\          &Profile  &Telescope \\
                &(2,4000,000+) &(\AA)         &         &   \\     
\hline\hline
02/06/2001      &52063.5       &-50.00$\pm$1.71    &SP   &SKO     \\
18/07/2002      &52474.5       &-34.82$\pm$0.80    &SP   &SKO     \\
11/09/2002      &52529.5       &-49.02$\pm$2.72    &SP   &SKO     \\
09/06/2003      &52800.5       &-38.69$\pm$3.50    &SP   &SKO     \\
24/06/2004      &53181.5       &-38.50$\pm$5.79    &SP   &SKO     \\
08/07/2004      &53195.5       &-35.80$\pm$3.47    &SP   &SKO     \\
\hline\hline
\end{tabular}
\end{center}
\end{table}

\clearpage

\begin{table}
\caption{Results for KS\,1947+300}
\label{ks1947}
\begin{center}
\begin{tabular}{l@{~}c@{~}c@{~~~~~~}l@{~~~~}l}
\hline  \hline
Date            &JD            &\ew\          &Profile  &Telescope \\
                &(2,4000,000+) &(\AA)         &         & \\     
\hline\hline
30/05/2001      &52060.5       &-18.58$\pm$0.43    &SP  &SKO      \\
01/06/2001      &52062.5       &-16.88$\pm$0.49    &SP  &SKO     \\
07/08/2001      &52129.5       &-16.28$\pm$0.47    &SP  &SKO     \\
09/08/2001      &52131.5       &-17.99$\pm$0.56    &SP  &SKO     \\
12/09/2001      &52165.5       &-15.99$\pm$0.28    &SP  &SKO     \\
08/10/2001      &52191.5       &-17.37$\pm$0.47    &SP  &SKO     \\
\hline\hline
\end{tabular}
\end{center}
\end{table}

\clearpage

\begin{table}
\caption{Results for GRO\,J2058+42}
\label{gro2058}
\begin{center}
\begin{tabular}{l@{~}c@{~}c@{~~~~~~}l@{~~~~}l}
\hline  \hline
Date            &JD            &\ew\          &Profile  &Telescope \\
                &(2,4000,000+) &(\AA)         &         &  \\     
\hline\hline
25/06/2004      &2453182.5       &-4.16$\pm$0.86    &DP    &SKO   \\
04/07/2004      &2453191.5       &-4.61$\pm$0.61    &DP    &WHT    \\
06/07/2004      &2453193.5       &-4.68$\pm$0.72    &DP    &SKO    \\
25/08/2004      &2453243.5       &-4.25$\pm$0.38    &DP    &SKO    \\
13/09/2004      &2453262.5       &-5.24$\pm$0.60    &DP    &SKO    \\
03/10/2004      &2453282.5       &-4.42$\pm$0.37    &DP    &SKO    \\
\hline\hline
\end{tabular}
\end{center}
\end{table}

\clearpage

\begin{table}
\caption{Results for SAX\,J2103.5+4545}
\label{sax2103}
\begin{center}
\begin{tabular}{l@{~}c@{~}c@{~~~~~~}l@{~~~~}l}
\hline  \hline
Date            &JD            &\ew\          &Profile  &Telescope \\
                &(2,4000,000+) &(\AA)         &         &      \\     
\hline\hline
01/08/2003      &52853.5       &-1.33$\pm$0.51    &DP   &SKO     \\
17/08/2003      &52869.5       &-1.28$\pm$0.45    &DP   &WHT     \\
14/09/2003      &52897.5       &+3.32$\pm$1.03    &AB   &WHT     \\
06/10/2003      &52919.5       &+2.80$\pm$0.47    &AB   &SKO     \\
08/10/2003      &52921.5       &+2.27$\pm$0.56    &AB   &SKO     \\
23/05/2004      &53149.5       &+3.13$\pm$0.29    &AB   &SKO     \\
\hline\hline
\end{tabular}
\end{center}
\end{table}

\begin{table}
\caption{Results for SAX\,J2103.5+4545 (cont.)}
\label{sax2103c}
\begin{center}
\begin{tabular}{l@{~}c@{~}c@{~~~~~~}l@{~~~~}l}
\hline  \hline
Date            &JD            &\ew\          &Profile  &Telescope \\
                &(2,4000,000+) &(\AA)         &         &      \\     
\hline\hline
30/09/2010      &55470.5       &-0.88$\pm$0.29    &SH   &SKO     \\
14/10/2010      &55484.5       &-0.96$\pm$0.71    &SH   &SKO     \\
20/08/2011      &55794.5       &2.76 $\pm$0.22    &AB   &SKO     \\
08/09/2011      &55813.5       &2.05 $\pm$0.64    &AB   &SKO*    \\
06/06/2012      &56085.5       &-2.73$\pm$1.10    &DP   &SKO     \\
25/08/2012      &56165.5       &-3.93$\pm$0.28    &DP   &SKO     \\
06/09/2012      &56177.5       &-5.21$\pm$0.38    &DP   &SKO     \\
13/09/2012      &56184.5       &-5.56$\pm$0.61    &DP   &SKO     \\
31/07/2013      &56505.5       &+1.60$\pm$0.29    &AB   &SKO     \\
30/08/2013      &56535.5       &+2.31$\pm$0.30    &AB   &SKO     \\
18/10/2013      &56584.5       &+2.77$\pm$0.23    &AB   &SKO     \\
03/11/2013      &56600.5       &+2.43$\pm$0.44    &AB   &FLWO    \\
07/12/2013      &56634.5       &+2.56$\pm$0.29    &AB   &FLWO    \\
06/01/2014      &56664.5       &+1.26$\pm$0.28    &SH   &FLWO    \\
05/08/2014      &56875.5       &-1.41$\pm$0.22    &DP   &SKO     \\
12/10/2014      &56943.5       &-3.76$\pm$0.12    &DP   &SKO     \\
\hline\hline
\end{tabular}
\end{center}
\end{table}

\clearpage

\begin{table}
\caption{Results for IGR\,J21343+4738}
\label{igr21343}
\begin{center}
\begin{tabular}{l@{~}c@{~}c@{~~~~~~}l@{~~~~}l}
\hline  \hline
Date            &JD            &\ew\          &Profile  &Telescope \\
                &(2,4000,000+) &(\AA)         &         & \\     
\hline\hline
30/07/2009      &55043.5       &+1.12$\pm$0.46    &SH   &SKO    \\
29/09/2009      &55104.5       &+1.18$\pm$0.43    &SH   &SKO    \\
28/08/2010      &55437.5       &+0.35$\pm$0.16    &SH   &SKO    \\
30/09/2010      &55470.5       &+0.32$\pm$0.27    &SH   &SKO    \\
20/08/2011      &55794.5       &-1.28$\pm$0.38    &SH   &SKO    \\
06/09/2011      &55811.5       &-2.19$\pm$0.72    &SH   &SKO    \\
24/08/2012      &56164.5       &-7.30$\pm$0.31    &SH   &SKO    \\
06/09/2012      &56177.5       &-8.36$\pm$0.63    &SH   &SKO    \\
13/09/2012      &56184.5       &-7.33$\pm$0.58    &SH   &SKO    \\
15/10/2012      &56216.5       &-5.21$\pm$0.33    &SH   &FLWO   \\
19/10/2012      &56220.5       &-5.95$\pm$0.48    &SH   &SKO    \\
26/12/2012      &56288.5       &-2.68$\pm$0.49    &SH   &WHT    \\
04/01/2013      &56297.5       &-1.88$\pm$0.30    &SH   &FLWO   \\
11/01/2013      &56304.5       &-0.72$\pm$0.48    &SH   &FLWO   \\
15/06/2013      &56459.5       &+1.68$\pm$0.35    &AB   &SKO    \\
05/07/2013      &56479.5       &+1.34$\pm$0.40    &AB   &FLWO   \\
31/07/2013      &56505.5       &+1.75$\pm$0.47    &AB   &SKO    \\
23/08/2013      &56528.5       &+1.93$\pm$0.52    &AB   &WHT    \\
18/10/2013      &56584.5       &+2.60$\pm$0.35    &AB   &SKO    \\
03/11/2013      &56600.5       &+2.61$\pm$0.11    &AB   &FLWO   \\
07/12/2013      &56634.5       &+2.93$\pm$0.10    &AB   &FLWO   \\
06/01/2014      &56664.5       &+2.43$\pm$0.20    &AB   &FLWO   \\
06/08/2014      &56876.5       &+2.36$\pm$0.94    &AB   &SKO    \\
19/08/2014      &56889.5       &+2.95$\pm$0.32    &AB   &SKO    \\
14/10/2014      &56945.5       &+2.56$\pm$0.31    &AB   &SKO    \\
19/10/2014      &56950.5       &+2.57$\pm$0.24    &AB   &FLWO   \\
\hline\hline
\end{tabular}
\end{center}
\end{table}

\clearpage

\begin{table}
\caption{Results for 4U\,2206+54}
\label{4u2206}
\begin{center}
\begin{tabular}{l@{~}c@{~}c@{~~~~~~}l@{~~~~}l}
\hline  \hline
Date            &JD            &\ew\          &Profile  &Telescope \\
                &(2,4000,000+) &(\AA)         &         &   \\     
\hline\hline
02/09/1990      &48137.5       &-2.48$\pm$0.65    &SH   &INT     \\
27/12/1990      &48253.5       &-3.85$\pm$0.55    &SH   &INT     \\
28/08/1991      &48497.5       &-3.03$\pm$0.78    &SH   &INT     \\
14/12/1991      &48605.5       &-2.46$\pm$0.95    &SH   &INT     \\
16/08/1992      &48851.5       &-3.85$\pm$0.52    &SH   &PAL     \\
17/08/1992      &48852.5       &-3.61$\pm$0.99    &SH   &PAL     \\
\hline\hline
\end{tabular}
\end{center}
\end{table}

\begin{table}
\caption{Results for 4U\,2206+54 (cont.)}
\label{4u2206c}
\begin{center}
\begin{tabular}{l@{~}c@{~}c@{~~~~~~}l@{~~~~}l}
\hline  \hline
Date            &JD            &\ew\          &Profile  &Telescope \\
                &(2,4000,000+) &(\AA)         &         &   \\     
\hline\hline
21/05/2007      &54242.5       &-4.15$\pm$0.88    &SH   &SKO     \\
16/05/2007      &54237.5       &-3.64$\pm$0.23    &SH   &FLWO*   \\
07/06/2007      &54259.5       &-3.78$\pm$0.21    &SH   &FLWO    \\
22/06/2007      &54274.5       &-4.29$\pm$0.39    &SH   &FLWO    \\
07/09/2007      &54351.5       &-4.61$\pm$0.23    &SH   &SKO     \\
11/09/2007      &54355.5       &-4.56$\pm$0.24    &SH   &SKO     \\
\hline\hline
\end{tabular}
\end{center}
\end{table}

\clearpage

\begin{table}
\caption{Results for SAX\,J2239.3+6116}
\label{sax2239}
\begin{center}
\begin{tabular}{l@{~}c@{~}c@{~~~~~~}l@{~~~~}l}
\hline  \hline
Date            &JD            &\ew\          &Profile  &Telescope \\
                &(2,4000,000+) &(\AA)         &         &\\     
\hline\hline
31/05/2001      &52061.558     &-9.05 $\pm$0.52   &DP    &SKO  \\
03/06/2001      &52064.479     &-9.59 $\pm$0.56   &DP    &SKO  \\
13/09/2001      &52166.417     &-12.25$\pm$1.13   &DP    &SKO  \\
08/10/2001      &52191.367     &-10.77$\pm$0.63   &DP    &SKO  \\
18/07/2002      &52474.481     &-10.05$\pm$0.61   &DP    &SKO  \\
08/10/2003      &52921.323     &-11.00$\pm$0.67   &DP    &SKO  \\
\hline\hline
\end{tabular}
\end{center}
\end{table}

\end{appendix}

\end{document}